\documentclass[9pt,twocolumn,twoside,lineno]{pnas-new1}
\templatetype{pnasresearcharticle} 

\title{Interaction instability of localization in quasiperiodic systems}

\author[a,1]{Marko \v Znidari\v c}
\author[a]{Marko Ljubotina} 

\affil[a]{Physics Department, Faculty of Mathematics and Physics, University of Ljubljana, Jadranska 19, SI-1000 Ljubljana, Slovenia}

\leadauthor{\v Znidari\v c} 

\significancestatement{
Understanding how small imperfections affect a system's dynamics is one of the central questions of theoretical physics, namely, do properties change in a smooth way, such that small perturbation leads to small changes, or do they change discontinuously. Localization in disordered many-particle quantum systems has been shown to be stable to interactions. On a single-particle level one can also achieve localizaton with a quasiperiodic potential -- a system without disorder but with rich properties. It is believed that localization in disordered as well as in quasiperiodic potentials behaves in essentially the same way, in particular, that both are stable against interactions. We show that this is not so. A quasiperiodic localized system discontinuosly changes from localization to diffusion upon introducing interactions.
}

\authorcontributions{M.~\v Z. designed research, performed Lindblad and single-particle analysis and wrote the paper, M.~L. performed unitary simulations.}
\authordeclaration{The authors declare no conflict of interest.}
\correspondingauthor{\textsuperscript{1}To whom correspondence should be addressed. E-mail: znidaric\makeatletter @\makeatother fmf.uni-lj.si}

\keywords{stability $|$ many-body localization $|$ disorder $|$ transport $|$ 1D quasicrystal} 

\begin{abstract}
  Integrable models form pillars of theoretical physics because they allow for full analytical understanding. Despite being rare, many realistic systems can be described by models that are close to integrable. Therefore, an important question is how small perturbations influence the behavior of solvable models. This is particularly true for many-body interacting quantum systems where no general theorems about their stability are known. Here, we show that no such theorem can exist by providing an explicit example of a one-dimensional many-body system in a quasiperiodic potential whose transport properties discontinuously change from localization to diffusion upon switching on interaction. This demonstrates an inherent instability of a possible many-body localization in a quasiperiodic potential at small interactions. We also show how the transport properties can be strongly modified by engineering potential at only a few lattice sites.  
\end{abstract}

\dates{This manuscript was compiled on \today}
\doi{\url{www.pnas.org/cgi/doi/10.1073/pnas.1800589115}}

\newcommand{\sx}[1]{\sigma^{\rm x}_{#1}}
\newcommand{\sy}[1]{\sigma^{\rm y}_{#1}}
\newcommand{\sz}[1]{\sigma^{\rm z}_{#1}}
\def\ii{{\rm i}}

\newcommand{\braket}[2]{\langle{#1}|{#2}\rangle}
\newcommand{\bracket}[3]{\langle{#1}|{#2}|{#3}\rangle}
\newcommand{\tr}{\mathrm{tr}}
\newcommand{\ave}[1]{\langle{#1}\rangle}

\def\tit#1{{\em #1},}
\def\etal#1{ {\em et al.}}

\usepackage{bm,bbm,cuted}

\begin{document}

\verticaladjustment{-2pt}

\maketitle
\thispagestyle{firststyle}
\ifthenelse{\boolean{shortarticle}}{\ifthenelse{\boolean{singlecolumn}}{\abscontentformatted}{\abscontent}}{}


\dropcap{S}cience tries to describe nature in the simplest possible terms. Models that can be solved exactly, for instance integrable systems, play a special role. Although generic systems are not integrable, physicists have developed a plethora of methods that successfully describe phenomena in terms of slightly perturbed integrable models. An important question that we address is how stable integrable systems are to perturbations.

One of the finest results in classical mechanics is the Kolmogorov-Arnold-Moser (KAM) theorem~\cite{Gutzwiller}. It shows that in a finite-dimensional classical system integrability breaks smoothly, that is, upon small perturbation most orbits retain their integrable quasiperiodic character and only few become irregular. Even though irregular orbits can form a connected ergodic component, the so-called Arnold web~\cite{Arnold}, their measure goes to zero as the perturbation strength decreases. Another important question is that of quantum localization. For non-interacting electrons in a periodic potential one can use the Bloch (Floquet) theorem to show that transport is ballistic (zero resistance), corresponding to extended eigenstates. It was believed that such systems would become diffusive in the presence of disorder, so it came as quite a surprise when Anderson showed~\cite{Anderson} that such ballistic transport is completely unstable against disorder; disorder can cause an immediate change from an ideal conductor to an ideal insulator -- the so-called Anderson localization. The KAM theorem of classical mechanics and quantum localization both use the same ``KAM techniques'' in their proofs~\cite{Pastur77,Frohlich86}, namely, dealing with a small denominator problem of perturbation series.

For systems with many particles, i.e., in the thermodynamic limit (TDL), things are less clear. A prevailing opinion is that general integrable systems are not stable, except perhaps localized ones, for which one can in some cases show that small nonlinearity e.g. preserves some localized orbits~\cite{Frohlich86b} or causes very slow transport~\cite{Basko11}. For quantum systems, even finite dimensional ones, no ``quantum KAM theorem'' is known, and due to an inherent ``discreetness'' of quantum mechanics even the very definition of integrability is not so clear-cut~\cite{Caux}. While traditional criteria from quantum chaos~\cite{Gutzwiller}, for instance the distribution of nearest eigenenergy spacings might suggest that the change is not smooth~\cite{Santos}, one has to note that these smallest energy scales in the TDL describe exponentially large (unphysical) times. A more relevant question is that of dynamics on a physically accessible time scale. 

As a dynamical criterion by which we judge the smoothness we study transport~\cite{Joel} in a one-dimensional system with a quasiperiodic potential~\cite{Satija}. Transport of interacting many-particle systems is of obvious high interest and has been discussed ever since the discovery of single-particle localization~\cite{Anderson}. Due to the difficulty, results were few~\cite{Fleishman80,Aronov85}, often focusing on a two-particle~\cite{Dima94,Flach12}, or single-particle problems with nonlinearity (that is supposed to describe interaction in an effective mean-field way)~\cite{Dima08,Flach09,Larcher12}. For the two-particle case results for small interactions and random~\cite{Dima94,Krimer11} as well as for (noncritical) quasiperiodic potential~\cite{Flach12} show that localization is preserved (at larger interactions two particles though can delocalize~\cite{Flach12}), while nonlinearities induce subdiffusive (or slower) transport. More recently, many-body localization (MBL), that is localization of many particles in the presence of interactions, has gained increased attention, for review see~\cite{rew,rew3}. MBL has been proven for small interactions in a random system~\cite{Imbrie}. Furthermore, many experiments probing non-interacting localization in fact employ a quasiperiodic disorder~\cite{Roati08,Lahini09} so it is important to understand how quasiperiodic disorder modifies localization properties. An important property of deterministic quasiperiodic potential is that there are no (rare) local configurations of small or large potential that could influence dynamical properties~\cite{Roeck17}. Quasiperiodic systems have been probed experimentally~\cite{Modugno11}, including the first demonstration of MBL~\cite{exper} (of certain degrees of freedom~\cite{Prelovsek16}). Experiments with quasiperiodic systems are by now well controlled~\cite{Luschen17,Luschen17b} and can even be performed in 2D~\cite{Bordia17}. A number of recent theoretical studies discussed MBL~\cite{Vadim13,Michal14,Vieri15,Varma15,Khemani17,Vieri17,Bera17,Lee17,Pixley17,Naldesi} or localization~\cite{Chandran17,Sarang17,Monthus17} in the cosine quasiperiodic potential. A common conclusion from all these few-particle as well as MBL studies seems to be that systems with quasiperiodic and random potential behave rather similarly (apart from a possibly different universality class of the transition point~\cite{Khemani17}). In particular, a noninteracting quasiperiodic localized system will behave smoothly as one adds interactions. This would be in-line with mathematical reasoning that a point spectrum (localization) is stable. We show that the situation is, in fact, the opposite.

\begin{figure}[htb]
\centerline{
\includegraphics[width=\linewidth]{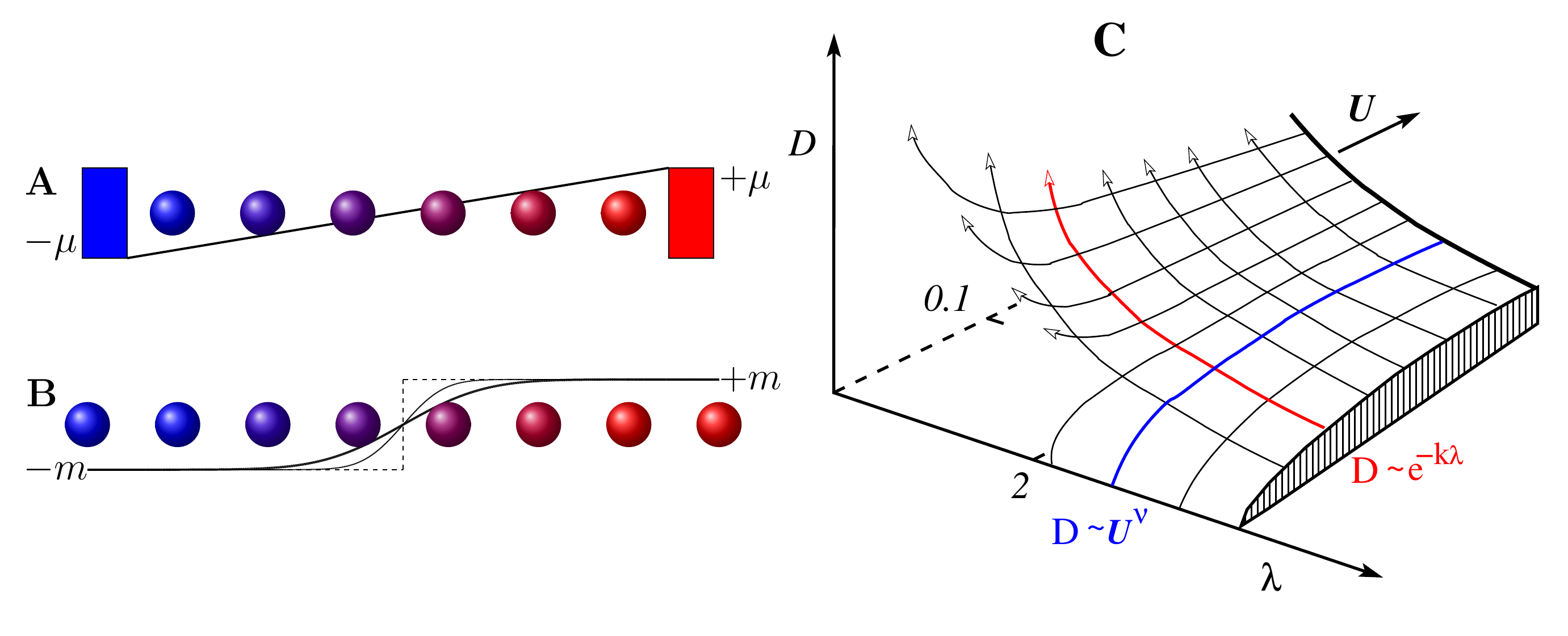}
}
\caption{Driven Lindblad setting (A) and unitary evolution (B) used in the study. Schematic diagram (C) of the diffusion constant $D$ on interaction $U$ and potential amplitude $\lambda$.}
\label{fig:shema}
\end{figure}

By studying transport properties of a one-dimensional interacting system in the presence of a quasiperiodic potential at infinite (high) temperature and half-filling we clearly demonstrate that the noninteracting localization discontinuously breaks down to diffusion for infinitesimal interactions (Fig.~\ref{fig:shema}). This surprising fragility of localization in a quasiperiodic potential seems to be due to long-range correlations (and resonances) in the single-particle spectrum, and must be contrasted with a continuous behavior for an uncorrelated potential. The result has several strong implications: (i) it shows that there can not exist a KAM-like quantum smoothness theorem that would hold in general for localized systems -- depending on the disorder type one can have smooth or non-smooth behavior; (ii) a possible MBL in quasiperiodic systems at small interactions is likely always unstable, again in contrast to the Anderson model which is stable~\cite{Imbrie}; (iii) by manipulating potential only at a few sites we can significantly affect the transport, opening the door to transport engineering.

\section*{The Model} 

We are going to study magnetization transport in the interacting Aubry-Andr\' e-Harper (AAH) model~\cite{Harper,AubryAndre},
\begin{equation}
  H=\sum_{j=1}^{L-1}(\sx{j} \sx{j+1} +\sy{j}\sy{j+1}+U \sz{j}\sz{j+1})+\sum_{j=1}^L h_j \, \sigma_j^{\rm z},
  \label{eq:H}
\end{equation}
with $h_j=\lambda\cos{(2\pi \beta j+\phi)}$ and $\sigma_j^\alpha$ a Pauli matrix. Using the Jordan-Wigner transformation Eq.~(\ref{eq:H}) is equivalent to spinless fermions with the interaction strength being given by $U$, while magnetization transport is equal to the particle transport. For irrational $\beta$ (we use $\beta=(\sqrt{5}-1)/2$) and without interaction ($U=0$) the model exhibits a transition from ballistic transport for $\lambda<2$ to full localization for $\lambda>2$~\cite{AubryAndre}, see~\cite{Damanik14} for rigorous results. The question we address is how this changes for $U \neq 0$.

Note that it is expected that a ballistic (integrable) system immediately breaks down to diffusion upon breaking its integrability. This is suggested by Fermi's golden rule: perturbation matrix elements are nonzero for extended states, leading to nonzero scattering and diffusion, as well as by explicit results, for instance, numerical studies of a gapless anisotropic Heisenberg chain in the presence of an extra coupling~\cite{Fabian03}, or staggered~\cite{Moore13} or random fields~\cite{Znidaric16} all show diffusion. Similarly~\cite{foot1}, exact results for free particles in the presence of a dissipative spin-conserving dephasing~\cite{Znidaric10} are also diffusive. In line with these expectations we also observe that the AAH model becomes diffusive for $\lambda<2$ as soon as $U \neq 0$. The more interesting case is the stability of a localized phase, $\lambda>2$, and this is the case we mostly focus on.

Transport is studied using two different settings (Fig.~\ref{fig:shema}). One is a true nonequilibrium steady state (NESS) situation, akin to what an experimentalist would do to measure conductivity. Boundary spins at each end of the chain are coupled to an effective bath that tries to induce an imbalance of magnetization between the left and the right end, causing a nonzero magnetization current. In technical terms, we use the Lindblad equation~\cite{GKS,Lindblad} whose solution gives the NESS density matrix $\rho_\infty$ to which any initial state converges after long time. We choose a Lindblad driving (see SI Appendix) that asymptotically induces small magnetization $\ave{\sz{1}} \approx +\mu$ at the left end and $\ave{\sz{L}}\approx-\mu$ at the right (typically $\mu =0.01$). Once we obtain the full many-body $\rho_\infty$ any observable in the NESS can be evaluated. In particular, we study the expectation value of the magnetization current operator, $j := \tr{[2(\sx{k} \sy{k+1}-\sy{k} \sx{k+1})\rho_\infty]}$, which is due to a continuity equation independent of the site index $k$. Provided one has diffusive transport, the current will (in the TDL) obey a phenomenological (Fourier's~\cite{Joel}) law
\begin{equation}
  j = -D \frac{\ave{\sz{L}}-\ave{\sz{1}}}{L},
  \label{eq:F}
\end{equation}
where $D$ denotes the diffusion constant. More generally, fixing the difference $\ave{\sz{L}}-\ave{\sz{1}}$, one can study the asymptotic dependence of $j$ on length $L$ and identify a scaling exponent $\gamma$ characterizing the transport type, $j \sim 1/L^\gamma$: $\gamma=1$ being diffusive, $\gamma=0$ ballistic, while $\gamma>1$ or $\gamma<1$ indicate sub- or superdiffusive transport, respectively.

The second method we employ is a unitary evolution of an inhomogeneous initial state, studying how the variance increases with time. For reasons of numerical efficiency a good choice~\cite{Ours17} seems to be a weakly polarized domain wall density matrix $\rho(0)\sim\left(e^{m\,\sigma^z}\right)^{\otimes\frac{L}{2}}\otimes\left(e^{-m\,\sigma^z}\right)^{\otimes\frac{L}{2}}$, where we take $m \approx \frac{\pi}{1800}$. Evolving it for a time $t$, $\rho(t)=U(t) \rho(0) U(t)^\dagger$, $U(t)={\rm e}^{-\ii H t}$, we calculate the magnetization profile and, specifically, magnetization transported across the middle of the chain, $\Delta s:=\sum_{j=1}^{L/2} \ave{\sz{j}}-\sum_{j=L/2+1}^{L} \ave{\sz{j}}$. In case of diffusion one expect the asymptotic dependence $\Delta s \propto \sqrt{t}$, while for anomalous transport one defines a scaling exponent,
\begin{equation}
  \Delta s \propto t^{\alpha},
  \label{eq:Ds}
\end{equation}
with $\alpha=1$ denoting ballistic transport, while $\alpha>1/2$ or $\alpha<1/2$ denote super- or sub-diffusion, respectively. For diffusion and long times the magnetization profile is given by the error function, $\ave{\sz{k}} \approx m\, {\rm erf}(\frac{k}{\sqrt{4D t}})$, from which we extract $D$. Assuming a single-parameter scaling of time and distance, $x \sim t^\alpha$, which leads to a time-of-flight across the chain $\tau \sim L^{1/\alpha}$ and in turn to current at fixed density $j \sim L/\tau$, results in a relation $\alpha=\frac{1}{\gamma+1}$. We use time-dependent density-matrix renormalization group (tDMRG)~\cite{dmrg3} for both numerical methods, enabling us to study large chains of up to $L=800$ sites. The results obtained from the Lindblad and unitary evolutions agree.

\section*{Diffusion} 

\begin{figure}[tb]
\centerline{\includegraphics[width=\linewidth]{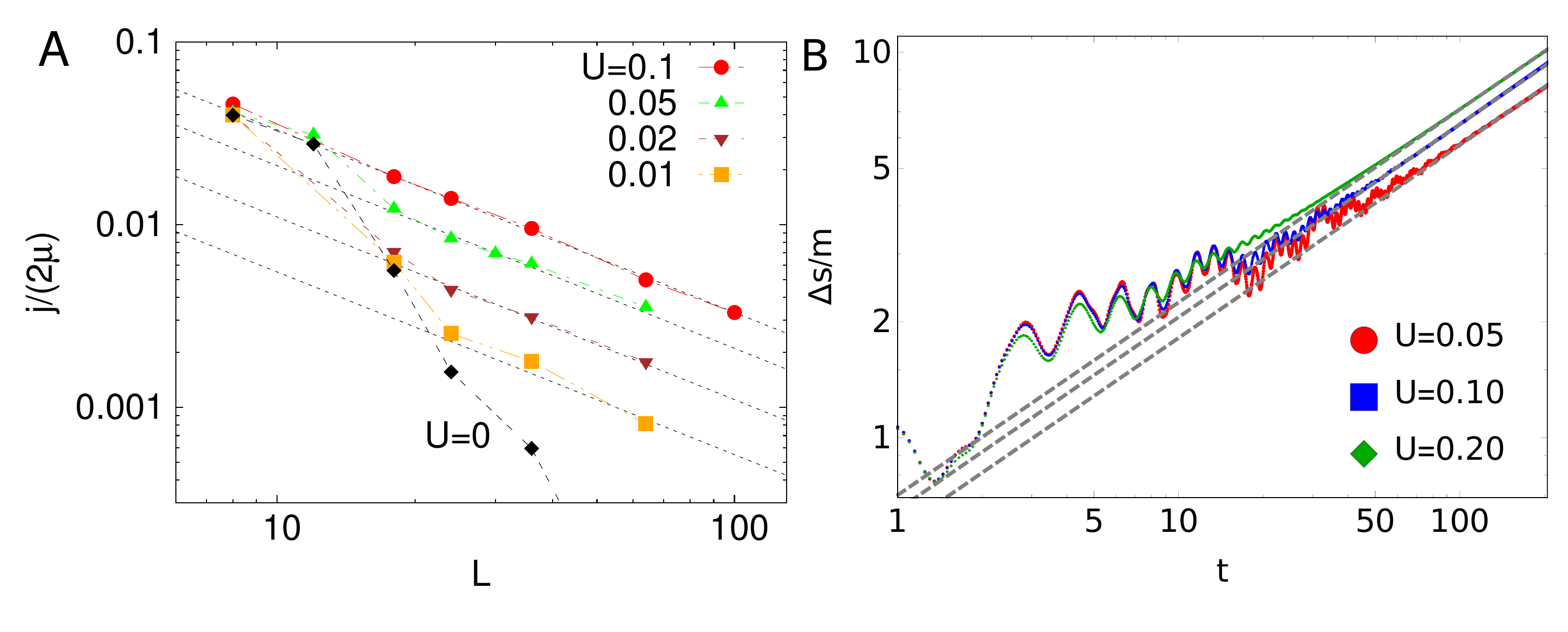}}
\caption{Diffusion for small interactions $U$ in the AAH model with $\lambda=2.2$. (A) NESS simulations. (B) Unitary domain wall spreading. Dashed lines denote diffusive scaling with increasingly smaller diffusion constant for decreasing $U$.}
\label{fig:diff}
\end{figure}

We first check the transport type for small interaction $U$ and $\lambda>2$, where all noninteracting states are exponentially localized. In the driven NESS setting we fix $\mu$, so that $\ave{\sz{1}}-\ave{\sz{L}} \asymp 2\mu$, and from the NESS current $j$ for different $L$ determine the scaling exponent $\gamma$. One can see in Fig.~\ref{fig:diff}a that, provided the system is large enough, one obtains diffusive $j \sim 1/L$ scaling regardless of how small the interaction is. What changes for small $U$ is that $j(L)$ follows the noninteracting exponentially small current (black dots in Fig.~\ref{fig:diff}a) up to increasingly larger $L$, when finally the asymptotic diffusion emerges. Similar results are also obtained for the unitary spreading shown in Fig.~\ref{fig:diff}b, where increasingly longer times are needed with decreasing $U$. We have an interesting situation where transport is diffusive at high energies while it is insulating at zero temperature~\cite{Vieri15,Vieri17}.

\begin{figure}[tb]
\centerline{\includegraphics[width=\linewidth]{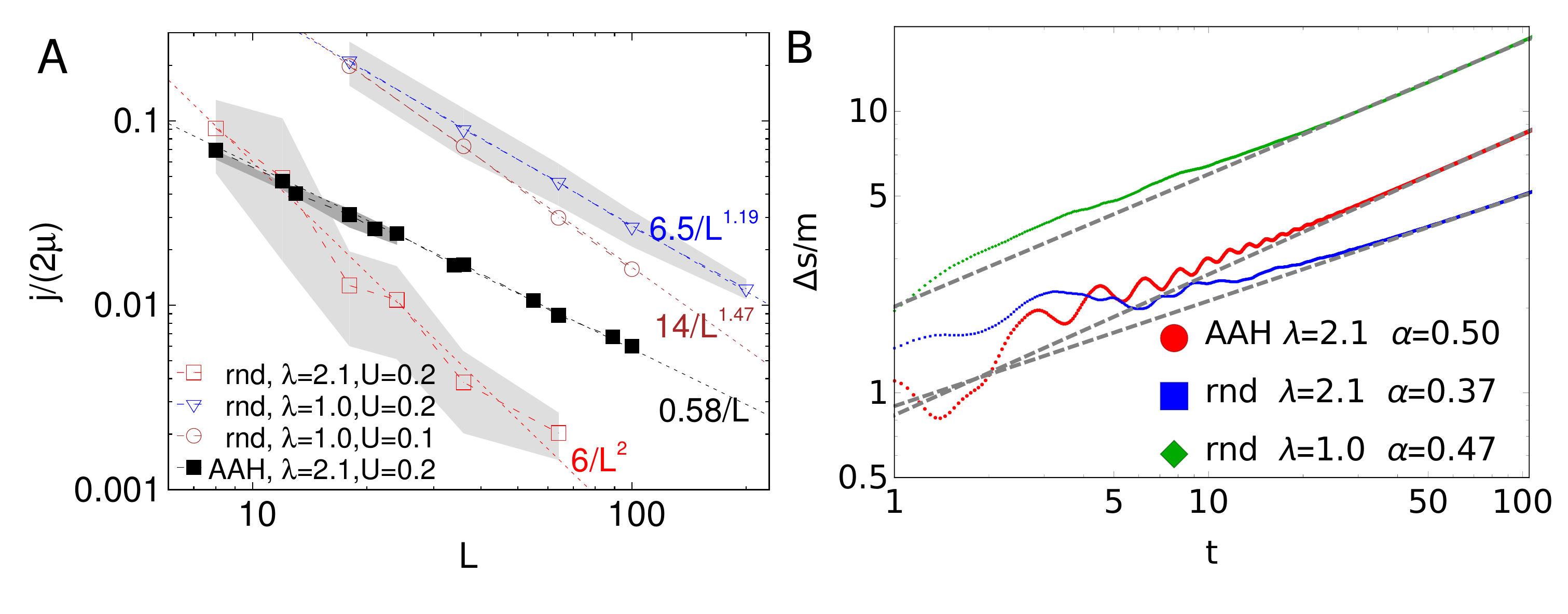}}
\caption{Comparing Anderson (rnd) and quasiperiodic potential (AAH) for $U=0.2$ (except brown circles in (A), where $U=0.1$). For the Anderson model localization breaks smoothly to subdiffusion. (A) Gray shading denotes standard deviation of the NESS current distribution obtained from 10 realizations of the phase $\phi$. (B) Unitary evolution and best fitting slope (\ref{eq:Ds}). The exponents $\gamma$ (in A) and $\alpha$ (in B) agree within numerical error via the relation $\alpha=\frac{1}{\gamma+1}$.}
\label{fig:cmp}
\end{figure}
\begin{figure*}[bt!]
\centerline{\includegraphics[width=\linewidth]{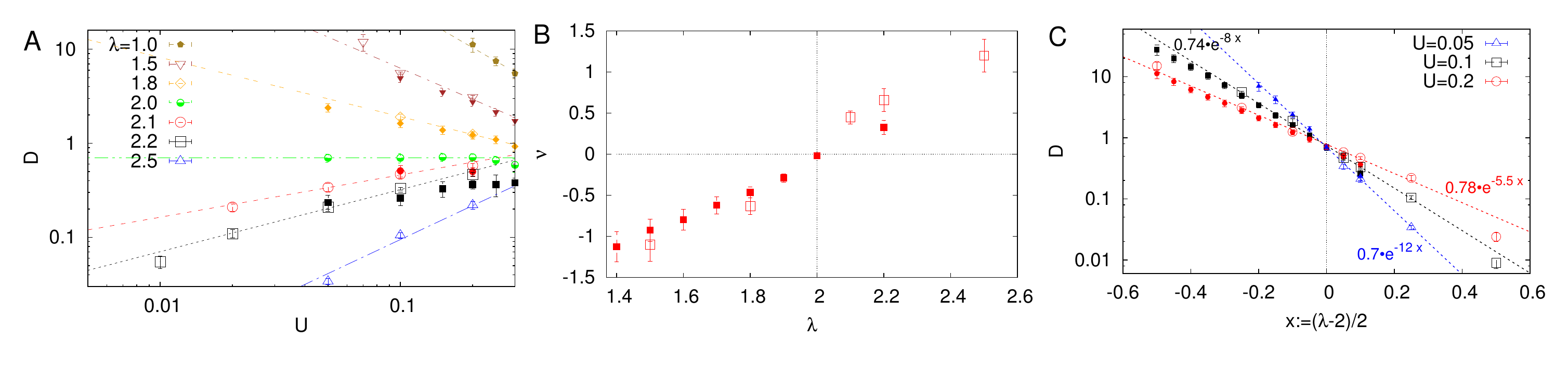}}
\caption{Diffusion constant dependence on interaction (A), scaling exponent $\nu$ on $\lambda$ (B), and $D$ on $\lambda$ (C). Empty symbols are obtained from NESS simulations, full from unitary evolution.}
\label{fig:D}
\end{figure*}
\begin{figure*}[t!]
\centerline{\includegraphics[width=0.91\linewidth]{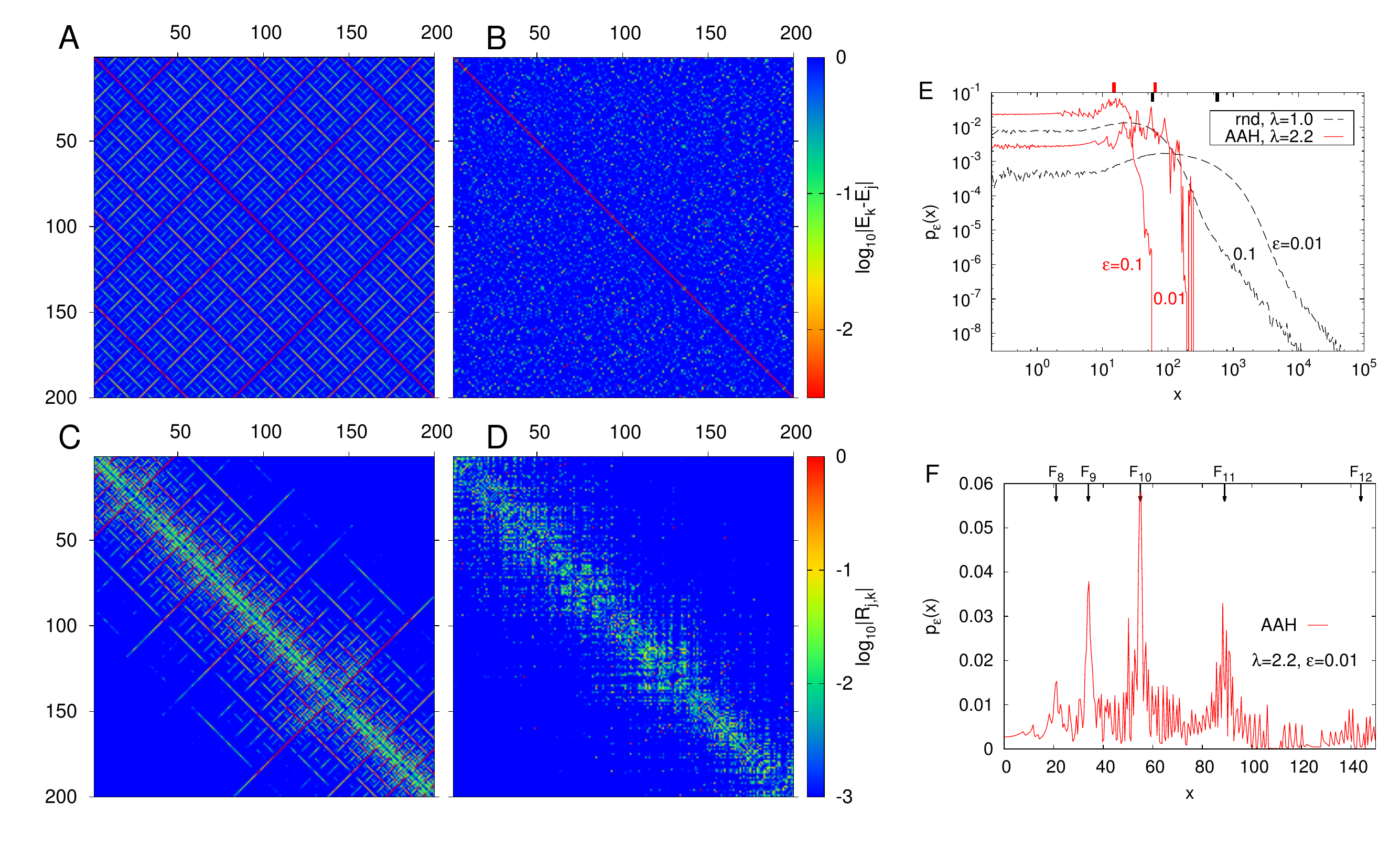}}
\caption{Single-particle resonances in the AAH model for $\lambda=2.2$ (A,C) and in the Anderson model for $\lambda=1.0$ (B,D). Two matrix indices in A,B,C,D label $200$ spatially consecutive eigenstates from the bulk. Top row (A,B) shows eigenenergy diferences, bottom (C,D) perturbation $R_{j,k}$ (defined in Eq.~(\ref{eq:R})). E and F show the distribution of distances to the nearest resonance with $|E_j-E_k|\le \varepsilon$ (F is an enlargement of the AAH data from E).}
\label{fig:res}
\end{figure*}
To stress the surprising nature of a discontinuous change from localization to diffusion for small $U$ we compare results to those for random potential at each site obtained by taking $h_j=\lambda \cos{(\phi_j)}$ with random independent phase $\phi_j$ at each lattice site, shortly called the Anderson model. Note that this does not change the distribution of the on-site potential~\cite{Khemani17}. Still, the localization length is different for the AAH and the Anderson model at the same $\lambda$ (see SI Appendix). To this end we also show the Anderson case at $\lambda=1.0$ where the non-interacting localization length is comparable to that of the AAH model at $\lambda=2.1$. Results shown in Fig.~\ref{fig:cmp} illustrate that in the Anderson case one gets subdiffusion~\cite{Znidaric16}, with the exponent $\gamma$ increasing as one decreases $U$. While the scaling exponents $\alpha$ are harder to distinguish (Fig.~\ref{fig:cmp}b) for the unitary evolution at relatively large $U=0.2$ ($\alpha = 0.50$ for the AAH and $\lambda=2.1$ vs. $\alpha \approx 0.47$ for the Anderson and $\lambda=1.0$), they are clearly different (Fig.~\ref{fig:cmp}a) for the Lindblad setting ($\gamma=1.0$ for the AAH and $\lambda=2.1$ vs. $\gamma \approx 1.19$ for the Anderson and $\lambda=1.0$). The difference only gets bigger if one decreases $U$ (or increases $\lambda$), e.g., for $U=0.1$ one has subdiffusive $\gamma\approx 1.47$ for the Anderson model at $\lambda=1.0$ (Fig.~\ref{fig:cmp}a), while for the AAH model at $\lambda=2.2$ one has diffusive $\gamma=1.0$ (Fig.~\ref{fig:diff}a). Sample to sample variation is considerably smaller in the AAH model (essentially no averaging over $\phi$ is needed) than in the Anderson model (Fig.~\ref{fig:cmp}a)~\cite{Khemani17}. Although both noninteracting models share at first sight similar single-particle localization, they behave completely differently in the presence of small interactions. We note that the instability of the AAH model has been noted before, namely, one can show that an infinitesimal modification of $h_j$ at a single site (preserving integrability) can change localization into an almost ballistic variance growth~\cite{Rio95}. A caveat is that these results are for specially constructed free systems of measure zero. Our results, on the other hand, demonstrate sensitivity for generic integrability breaking perturbation.

Confirming diffusion for small $U$, we study in detail the dependence of the diffusion constant $D$ on $U$ and $\lambda$. We find that $D$ goes to zero as one decreases $U$ for $\lambda>2$, while for $\lambda<2$ diffusion constant diverges (see SI Appendix). In both cases the dependence follows (Fig.~\ref{fig:D}a)
\begin{equation}
  D \approx U^\nu, \qquad U \ll 1.
  \label{eq:D}
\end{equation}
The scaling exponent $\nu$ is a smooth function of $\lambda$, being negative for $\lambda<2$ and positive for $\lambda>2$, see Fig.~\ref{fig:D}b. At fixed interaction $D(\lambda)$ is described rather well by an exponential function, Fig.~\ref{fig:D}c. At small $U$ and large $\lambda$ it gets exponentially small; if expressed in terms of a non-interacting single-particle localization length $\xi$ (see SI Appendix) the dependence would be $D \sim {\rm exp}(-4 {\rm e}^{1/2\xi}\ln\frac{1}{U})$ for $\lambda \gg 2$. It rapidly decays for small $\xi$ which has serious consequences for any numerical method that tries to probe that regime as it becomes exceedingly difficult to resolve this very small $D$ from other possible transport types (localization or subdiffusion). In our tDMRG simulations two hurdles prevent us to go to even larger $\lambda>3$: one is slow dynamics (i.e., small $j$) resulting in very long convergence times, another is increasingly large matrix-product dimensions required to get the NESS. While these obstacles prevent us to probe the regime close to a possible MBL transition at large $\lambda$, we do not see any signs of localization for $\lambda \le 3$ and small $U$ (see SI Appendix). Our results are thus not compatible with the MBL transition point smoothly connecting to $\lambda=2$ for $U \to 0$. The simplest conjecture is that at arbitrary fixed $\lambda$ transport is diffusive for sufficiently small interaction $U$ (numerically we can not exclude different behavior at large $\lambda$ when the single-particle localization length $\xi \ll 1$). It is not clear if rare-region arguments like in Ref.~\cite{Roeck17} could be used to explain our findings. Another interesting point that needs further attention is behavior for $\lambda=2$. We can see in Fig.~\ref{fig:D}c that the curves for $D(\lambda)$ (almost) cross at $\lambda=2$, indicating that $D$ is within our precision essentially independent of $U$ (see also Fig.~\ref{fig:D}a), and equal $D(\lambda=2,U=0.05) \approx 0.7$, approximately agreeing with $\approx 0.55$ obtained in Ref.~\cite{Vipin17,Manas17} for $U=0$ by looking at the spreading of an initial delta-packet (on times larger than our simulations for $U\neq 0$).

\section*{Single-particle correlations}
In the following we give an explanation why a quasiperiodic potential is so much different than a random one. Taking $U=0$ we calculate all single-particle eigenstates $\psi_k$ and eigenvalues $E_k$ of a long chain of $L = 10^6$ sites (see SI Appendix) for the AAH and the Anderson model. Spectral index $k$ is ordered according to eigenstate's center-of-mass location $c_k$, $c_k:=\sum_l l\,|\braket{l}{\psi_k}|^2$, so that $c_k \le c_{k+1}$. Looking at the single-particle eigenenergy resonances $E_k-E_j$ we notice (Fig.\ref{fig:res}) that in the AAH model they have a deterministic (fractal) spatial structure, while in the Anderson model they are random. More quantitatively, we determine for each $E_k$ the distance $x_k=c_l-c_k$ to the nearest resonance with an energy mismatch smaller than $\varepsilon$, $x_k={\rm min}_{l \neq k}(|c_l-c_k|; |E_k-E_l|\le \varepsilon)$. From all $10^6$ $x_k$ we determine the resonance distance distribution $p_\varepsilon(x)$, which is shown in Fig.~\ref{fig:res}e,f. We can see that in the AAH model these distances are much smaller than in the Anderson case (and have a sharp cut-off with no long tails), and are greatly enhanced at the Fibonacci numbers $x=F_n$ (Fibonacci $F_n$ give the best rational approximates to our $\beta\approx F_n/F_{n+1}$). This is also visible in Fig.~\ref{fig:res}a, where the main resonances are seen at $l \pm k=F_n$, corresponding to locations with similar local potential, $\cos{(2\pi \beta l+\phi)}-\cos{(2\pi \beta k+\phi)} \sim |\sin{(\pi\beta(l-k))}\sin{(\pi\beta(l+k)+\phi)}| \ll 1$, see also~\cite{Sarang17}. These same resonances are also reflected in the enhanced 2-particle matrix elements of the perturbation $V$
\begin{equation}
  R_{l,k}=\frac{\bracket{\psi_{l-1},\psi_l}{V}{\psi_{l-1},\psi_k}}{E_l-E_k},\quad V=\frac{1}{4} \sum_r \sz{r}\sz{r+1}, 
  \label{eq:R}
\end{equation}
that determines whether two single-particle eigenstates hybridize or not. In short, while the resonances are spatially random in the Anderson model, with some eigenstates without any resonant neighbors within the localization length, leading to subdiffusion, in the AAH model there is an enhanced probability for a nearby resonance at distance $\approx F_n$. We are at present not able to quantitatively explain our diffusion, e.g., the scaling exponent $\nu$ (\ref{eq:D}), just from $R_{l,k}$. It could be that a finite density of particles (a 2-particle problem still shows localization, see also Ref.~\cite{Flach12}) and higher order terms are needed for that.

\section*{Transport engineering}

\begin{figure}[htb]
\centerline{\includegraphics[width=\linewidth]{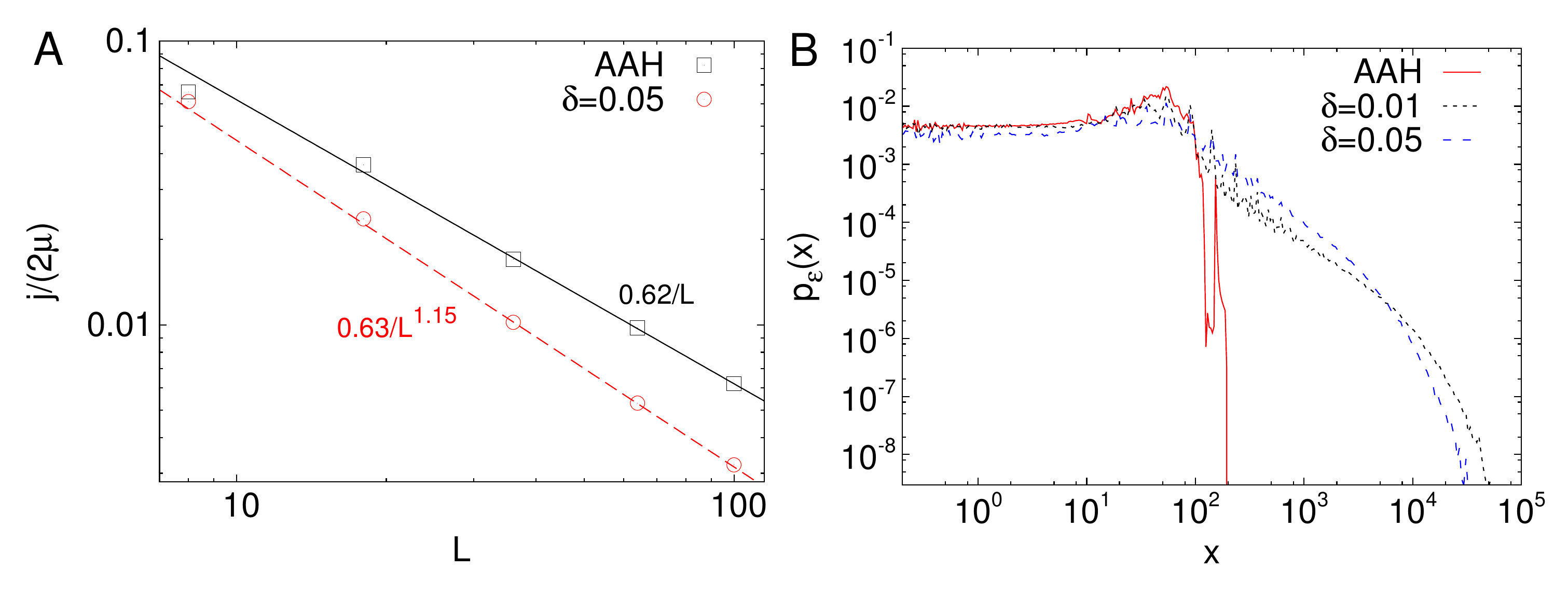}}
\caption{(A) Randomizing the phase at only few sites changes transport from diffusive to subdiffusive. (B) For small $\delta$ the distribution of nearest-resonance distances immediately gets a long tail, $\varepsilon=0.01$. All for $\lambda=2.1$, $U=0.2$.}
\label{fig:kick}
\end{figure}
Finally, we demonstrate that it is indeed the spatial ``Fibonacci weave'' correlations in $E_k$ (Fig.\ref{fig:res}a) due to the quasiperiodic potential that is crucial for the diffusion. To break these correlations we take the AAH model with a modified phase at some sites. Namely, at sites $j$ where $1-\cos{(2\pi \beta j+\phi)} < \delta$, we instead pick a random phase $\phi_j$ so that at that site $h_j=\lambda \cos{\phi_j}$. For small $\delta$ this still preserves the local cosine shape of the potential as we randomize only a few sites (for $\delta=0.05$ approximately every tenth site). Despite that correlations in $E_k$ break (see SI Appendix) and the transport immediately changes from diffusive to subdiffusive, Fig.~\ref{fig:kick}. Switching on interactions and modifying the potential at a few sites therefore enables one to alter the global transport type. It would be interesting to see how such changes influence transport in other free models with a correlated potential~\cite{felix12}. This suggests a mindset change from that of studying transport of particular models to one where one instead asks how can one engineer a specific transport type? For free models there are results that allow to construct a model with any anomalous transport~\cite{Mantica97,Guarneri99,Jitomirskaya07}.

\section*{Conclusions} 
We have shown that exponential localization in a quasiperiodic potential in the presence of small interactions at half-filling and high energy breaks down discontinuously to diffusion. This inherent instability, which is markedly different than the smooth behavior seen for a random potential, appears to be due to long-range correlations in the single-particle spectrum. They have implications for possible proofs of MBL in quasiperiodic systems, while high sensitivity of global transport to the potential at only a few sites opens the way to transport engineering, where one could significantly modify transport by small targeted changes to the potential. Our findings should be within reach of present day experiments. 

\acknow{We would like to thank T.~Prosen and V.~K.~Varma for comments on the manuscript. This work is supported by Grants No. J1-7279 and P1-0044 from the Slovenian Research Agency, and by Advanced grant of European Research Council (ERC) 694544 -- OMNES.}

\showacknow 

\subsection*{References}




\newpage
\clearpage

\begin{strip}
 {\Huge\sffamily\bfseries Supporting Information: Interaction instability of localization in quasiperiodic systems}
\end{strip}

\section*{M. \v Znidari\v c \& M. Ljubotina}

\setcounter{equation}{0}
\setcounter{figure}{0}
\setcounter{table}{0}
\makeatletter
\renewcommand{\theequation}{S\arabic{equation}}
\renewcommand{\thefigure}{S\arabic{figure}}

\section*{Numerical methods} 

\subsection*{Lindblad equation}

For the NESS setting we use a boundary driven Lindblad equation~\cite{GKS,Lindblad},
\begin{equation}
\frac{{{\rm d}}\rho}{{{\rm d}t}}=\ii [ \rho,H ]+ \sum_{k=1}^4 \left( [ L_k \rho,L_k^\dagger ]+[ L_k,\rho L_k^{\dagger} ] \right),
\label{eq:Lin}
\end{equation}
with Lindblad jump operators $L_k$ given by $L_1=\sqrt{1+\mu}\,\sigma^+_1, L_2= \sqrt{1-\mu}\, \sigma^-_1$ at the left end, and $L_3 =  \sqrt{1-\mu}\,\sigma^+_L, L_4= \sqrt{1+\mu}\, \sigma^-_L$ at the right end, $\sigma^\pm_k=(\sigma^{x}_k \pm {\rm i}\, \sigma^{y}_k)/2$. Such Lindblad equation has a unique steady state $\rho_\infty$ to which any initial state converges after sufficiently long relaxation time. The NESS $\rho_\infty$ is found by calculating the time evolution, $\rho(t) = {\rm e}^{{\cal L}t} \rho(0)$, Trotterizing the evolution into small time steps of $\Delta t=0.05$ and using a matrix product operator (MPO) ansatz with bond dimension $\chi$ and the tDMRG algorithm~\cite{dmrg3}. The algorithm that we use has been used in a number of works, see e.g. references cited in~\cite{Znidaric16}.

For $\mu=0$, which would correspond to an equilibrium driving (no imbalance between the left and right end), the steady state is simply $\rho_\infty \propto \mathbbm{1}$ -- an equilibrium state at infinite temperature. For small nonzero $\mu$, we typically use $\mu=0.01$, the NESS is still close to the identity, and therefore describes the nonequilibrium physics close to an infinite temperature. The decisive parameter for numerical efficiency is the required MPO bond dimension (typically $50-200$) and the relaxation time. We check the accuracy and estimate the error of the calculated NESS current by repeating calculations with progressively larger $\chi$ until a satisfactory convergence is reached. For large $\lambda$ (see Fig.~\ref{fig:lam30}), when the diffusion constant $D$ is small, the relaxation time can go up to $10^4$ for $\lambda=3$. Namely, the required relaxation time is at least of the order of $1/j$ (to reach the steady state magnetization has to be ``pumped'' out of the systems at its boundaries, which proceeds with ``speed'' $\sim j$). Because $D$ decreases exponentially for large $\lambda$ (Fig.~\ref{fig:D}c) this means that the relaxation time grows rapidly. Additionally, the required $\chi$ can become larger than $200$. One reason why typically larger $\chi$ are required is that in order to have some fixed final precision in the NESS the error per time step has to be smaller when relaxation time is longer. We also note that the MPO bond dimension of density operators is a priori not related to the entanglement of $\rho(t)$ and so in general smaller entanglement expected for large $\lambda$ does not necessarily translate to smaller $\chi$. All this unfortunately prevents us from studying large $\lambda$ close to a possible MBL transition point.

\subsection*{Unitary evolution}

In the unitary setting we evolve a weakly polarized domain wall initial state (see main text) with the von Neumann equation
\begin{equation}
\frac{{{\rm d}}\rho}{{{\rm d}t}}=\ii [ \rho,H ].
\label{eq:VonN}
\end{equation}
We then follow the spin and current at each site of the chain and obtain the transported magnetization, which allows us to determine the type of transport, as well as the diffusion constant (where it applies), which is determined by fitting an error function profile to the spin profiles at different times. Similarly to the Lindblad case we Trotterize the evolution into small time steps $\Delta t=0.01$ and use a matrix product ansatz in combination with tDMRG to perform the numerical simulation. For more details see \cite{Ours17}. Typically we used a spin chain of length $L=400$ or $L=800$, such that there are no edge effects at the maximal time.

\begin{figure}[t!]
\centering
\includegraphics[width=0.77\linewidth]{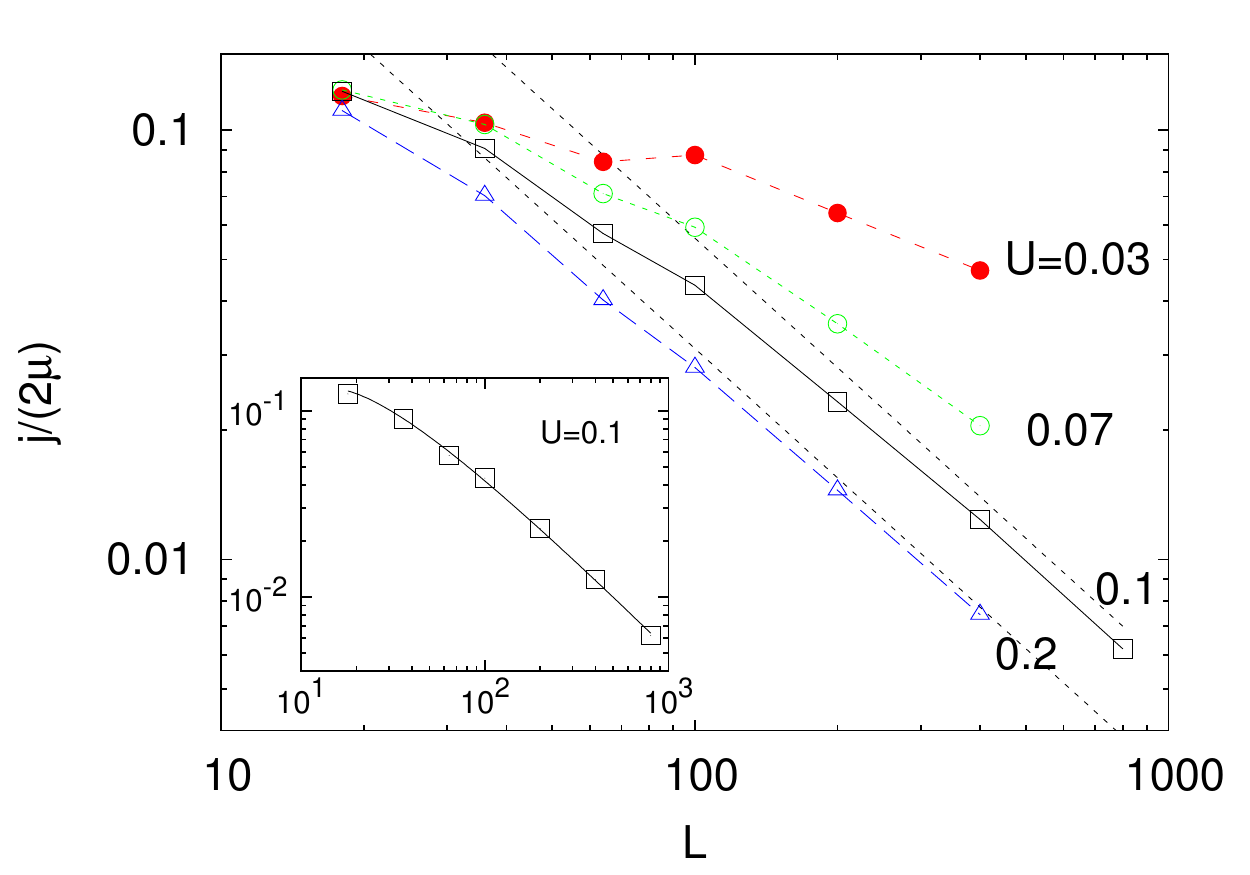}
\caption{NESS current for the AAH model and $\lambda=1.5$. Dotted lines denote diffusion. The inset show a best fit (full line) beyond the leading order for $U=0.1$, and is $j/(2\mu)=\frac{5.6}{L}(1-\frac{2.5}{L^{0.5}})$, meaning that $D=5.6$.}
\label{fig:lam15}
\end{figure}
\begin{figure}[t!]
\centering
\includegraphics[width=0.73\linewidth]{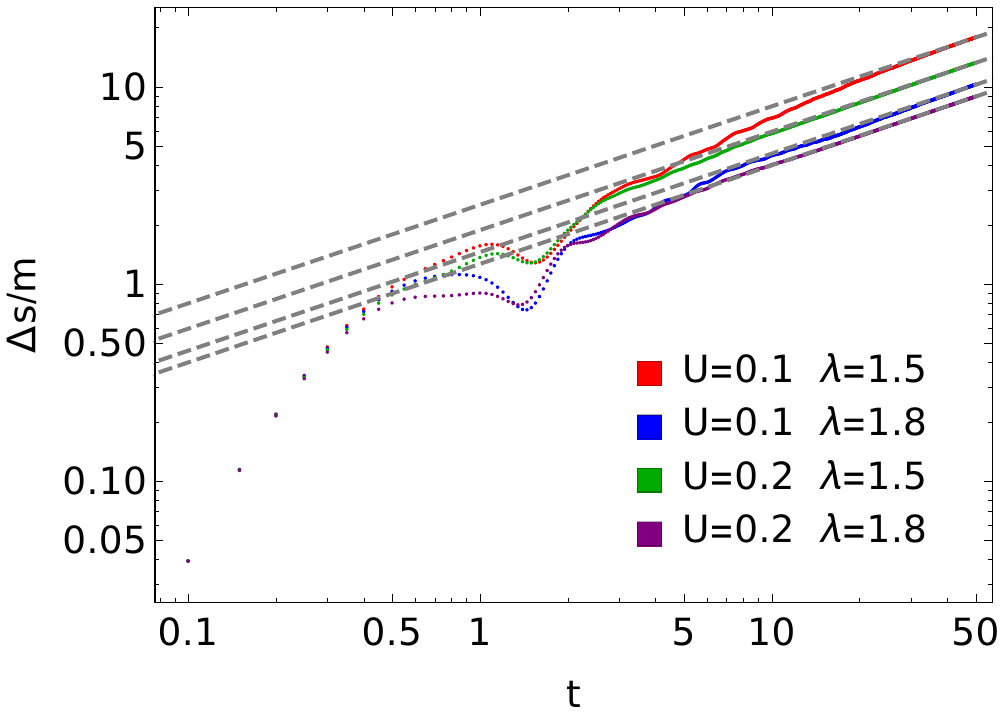}
\caption{Transferred magnetization in the unitary time evolution for the AAH model for $U=\{0.1,0.2\}$ and $\lambda=\{1.5,1.8\}$. The dashed lines denote the diffusive growth with a slope of $\alpha=\frac{1}{2}$. }
\label{fig:sfig2}
\end{figure}

In the case of diffusive transport this approach works well in the intermediate regime, where the diffusion constant is neither too small nor too large. Outside of this regime the primary difficulty becomes the time needed to pass from a transient phenomenon to the actual transport one would expect to see in the TDL. As usual with tDMRG increasing times usually come with increasing bond dimensions which quickly increases the simulation times beyond reasonable values. Additionally, in the case of a very small diffusion constant, the procedure with which we determine the diffusion constant becomes increasingly less accurate as the profiles become narrower. Again, this can only be alleviated by further increasing the times and thus allowing the profile to spread across a larger portion of the chain. On the up side, it could be possible to use shorter chains in this case, assuming of course that this has no impact on the time evolution, which can be difficult to prove. 

\section*{Additional data}

\subsection*{Diffusive regime}

Here we show additional data for $\lambda<2$, where transport expectedly becomes diffusive. 
In Fig.~\ref{fig:lam15} we can see that the asymptotic dependence of the NESS current is $j \sim 1/L$. For small $U$, where the scattering rate of perturbation (interaction $V$) is small, and the corresponding scattering length is large, one needs a large system size $L$ in order to reach the asymptotic diffusive regime. For instance, for $U=0.1$ we calculated the NESSs for $L \le 800$ (with bond dimensions $\chi=100$ we could reach less than $5 \%$ error in the NESS current). As one can see from the fit (the inset of Fig.~\ref{fig:lam15}) even at that size the subleading correction amounts to about $9 \%$. For even smaller $U=0.07$ and $U=0.03$ it is impossible to reach the asymptotic diffusive regime with our computational resources (hundreds of CPU cores and weeks of CPU time). The same also holds for the unitary case where the time one needs to evolve the system to, in order for it to relax to the diffusive state, typically diverges as $U$ approaches zero regardless of $\lambda$, except perhaps for $\lambda=2$ where $D$ changes little.
\begin{figure}[bt]
\centering
\includegraphics[width=0.77\linewidth]{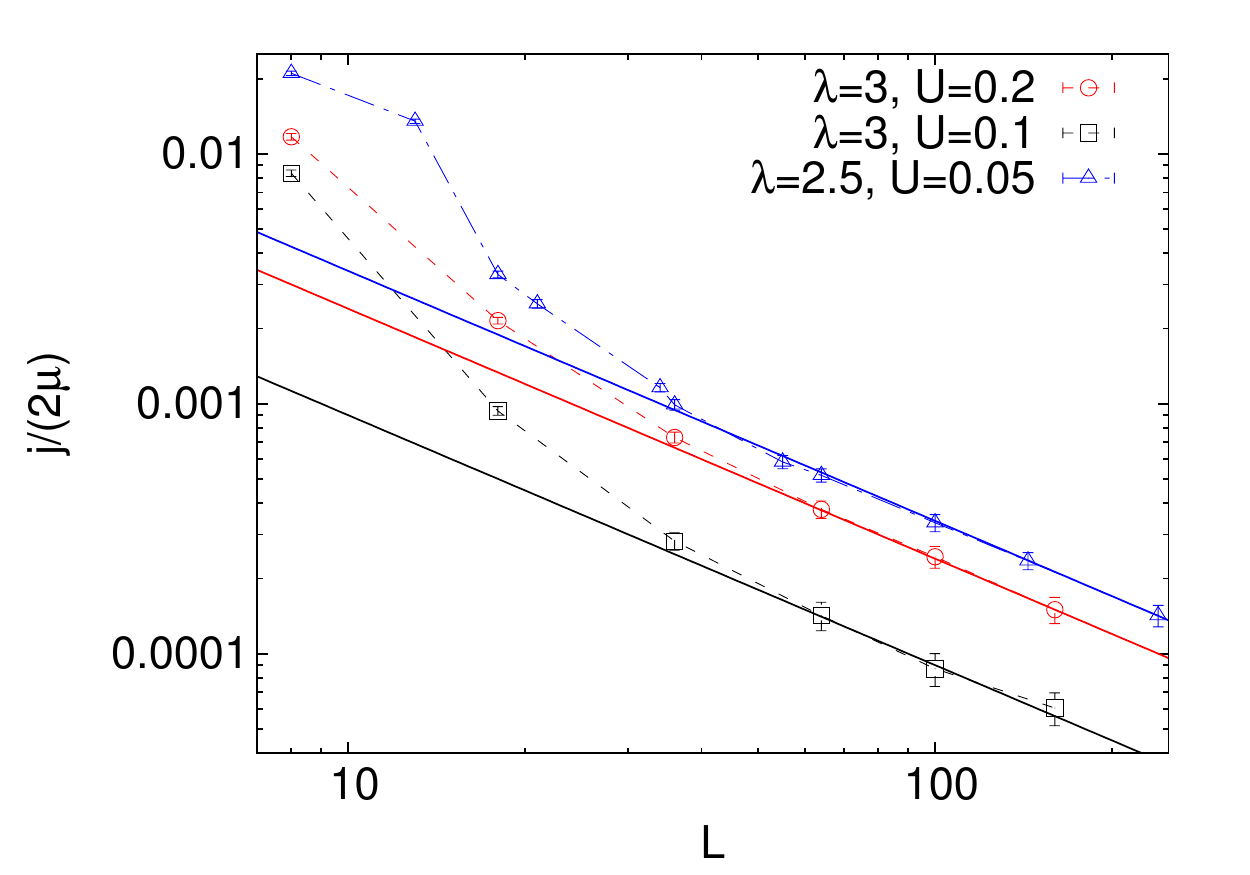}
\caption{NESS current for a single realization of the AAH model at $\lambda=3.0$ and $\lambda=2.5$. Numerics is very difficult here, and one also needs larger $L$ to reach the asymptotic behavior. Full lines are best fitting diffusive $j \sim D 2\mu/L$ dependence, while the error bars denote an error estimate due to finite MPO size $\chi$.}
\label{fig:lam30}
\end{figure}
\begin{figure}[bt]
\centering
\includegraphics[width=0.77\linewidth]{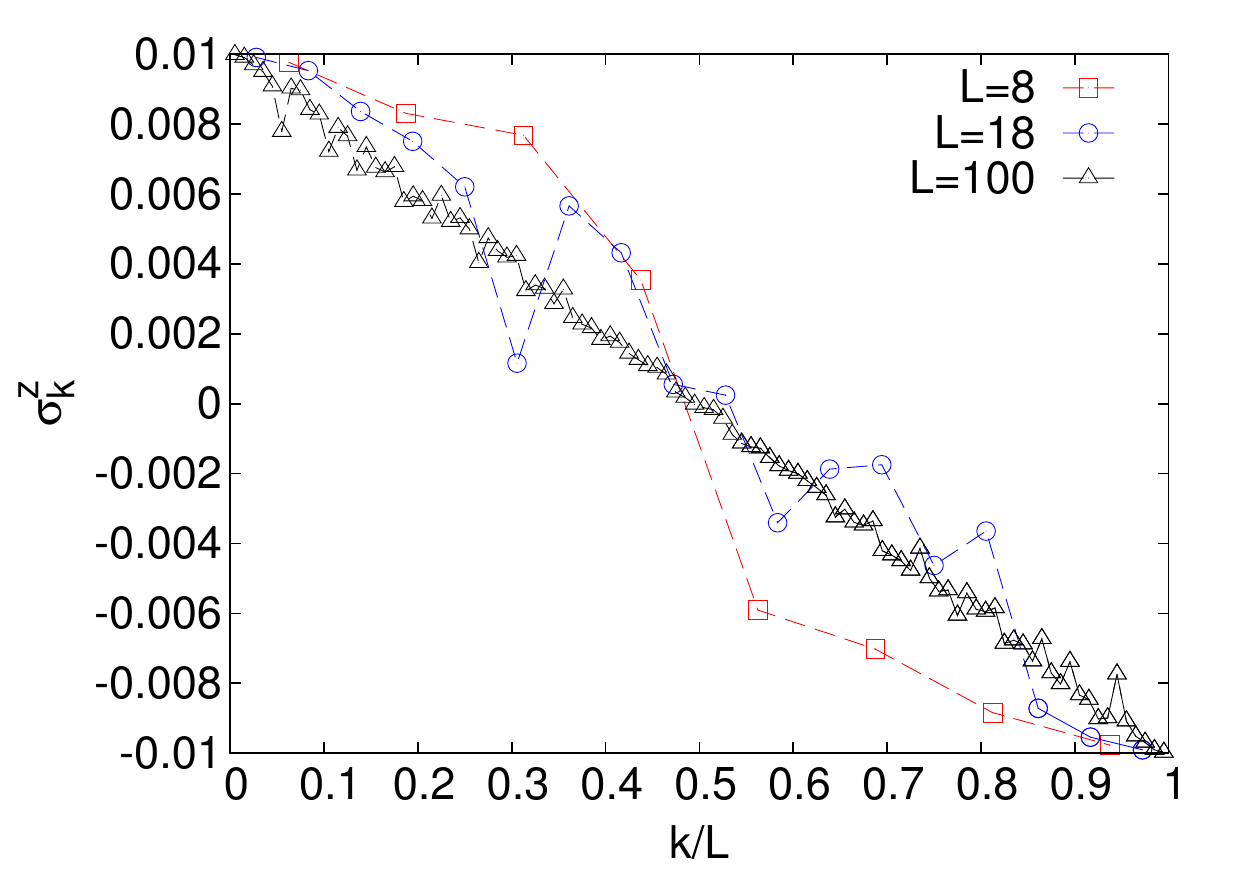}
\caption{NESS magnetization profile for one instance of the AAH model with $\lambda=2.2$ and $U=0.1$.}
\label{fig:profil22}
\end{figure}
\begin{figure}[th!]
\centering
\includegraphics[width=0.75\linewidth]{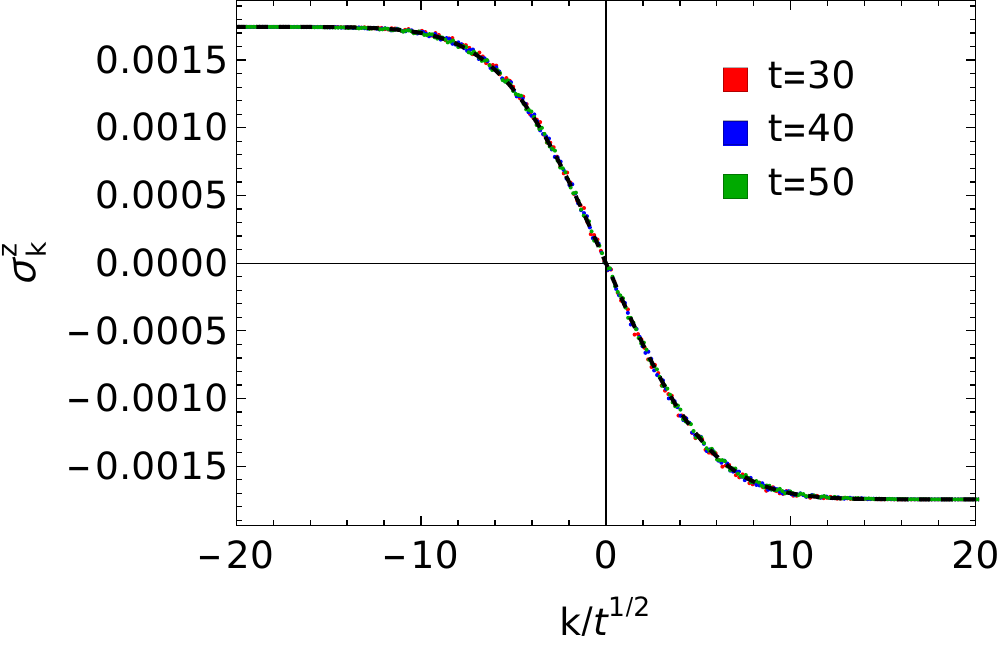}
\includegraphics[width=0.75\linewidth]{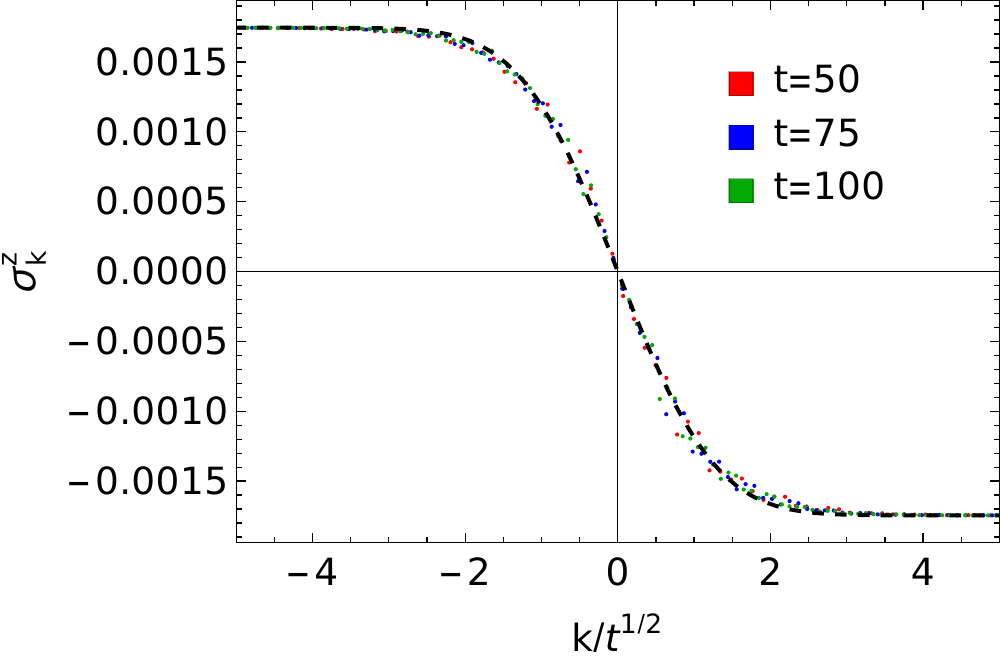}
\caption{Spin density profiles for the AAH model at different times and unitary evolution of an initial step-like density operator. One sees that the lines at different times collapse excellently onto one another. The dashed black line shows an error function fit to the data at the latest time. Top figure is for $\lambda=1$ with the best-fitting diffusion constant $D \approx 11$, bottom for $\lambda=2.1$ and $D\approx 0.5$, both at $U=0.2$.}
\label{fig:sfig3}
\end{figure}
\begin{figure}[bt]
\centering
\includegraphics[width=0.75\linewidth]{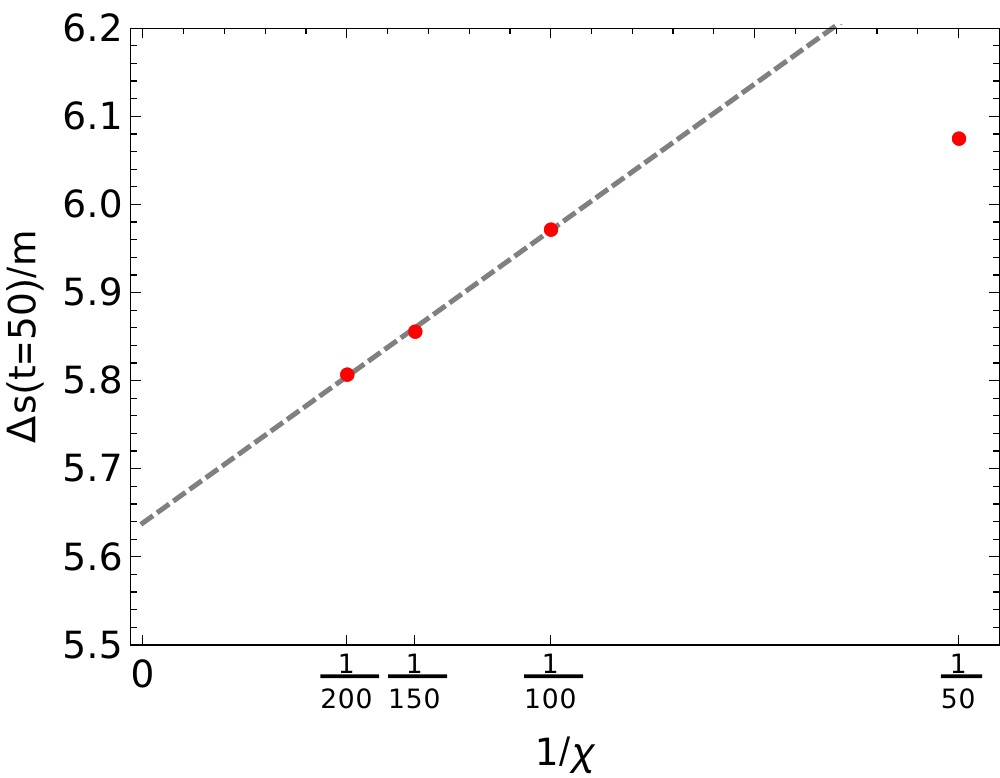}
\caption{Transferred magnetization at $t=50$ for different bond dimensions for the AAH model with $\lambda=2.1$ and $U=0.2$ and unitary evolution. The dashed line shows a fit which suggests a convergence to a final value of $5.64$ (at $\chi\to\infty$) which is $<3\%$ from the value at $\chi=200$. } 
\label{fig:sfig1}
\end{figure}
\begin{figure*}[tb!]
\centerline{\includegraphics[width=0.6\linewidth]{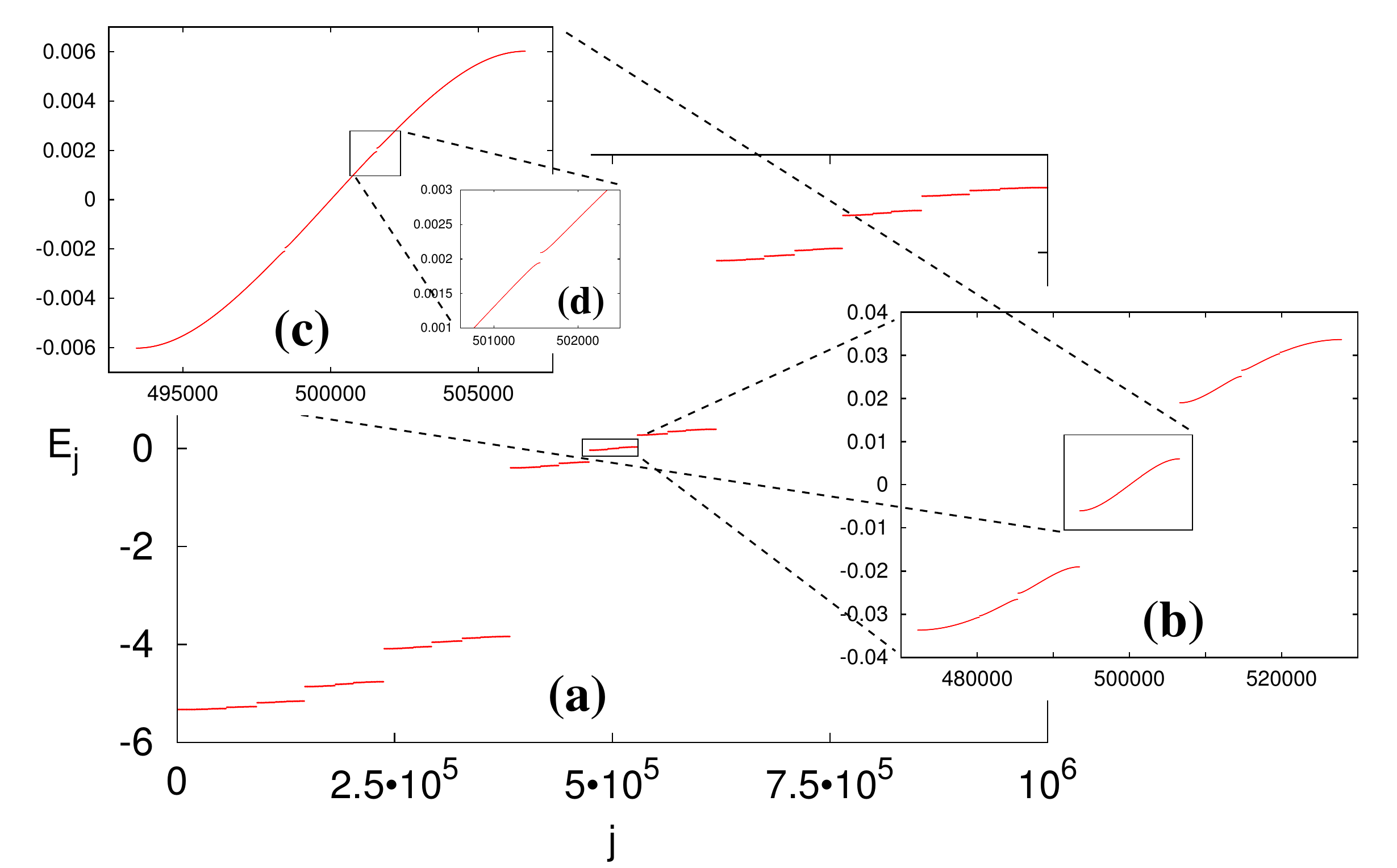}}
\caption{The single-particle spectrum of the AAH model ($U=0$) with $\lambda=2.1$ and $L=10^6$. Eigenvalues $E_j$ are ordered, with different zoom-ins illustrating the self-similar structure of the spectrum. On a global scale there are three main bands (level 0, frame (a), containing $L=10^6$ eigenenergies). Each of those is then split into an infinite series of sub-bands. For instance, frame (b) shows bands at level 2 (centered around $E=0$), consisting of about $5\cdot 10^4$ eigenstates, while the final zoom-in in frame (d) shows a tiny ``band-gap'' between the level 4 bands, each of which contains only $\sim 3000$ eigenstates (the number of eigenstates in a band at level $r$ is $L \cdot (\sqrt{5}-2)^r$ and for $r=5$ gets too small to resolve the next level gap, whose size shrinks exponentially with $r$).}
\label{fig:spekterAAH}
\end{figure*}

As already mentioned, NESS numerics gets harder for large $\lambda$. In Fig.~\ref{fig:lam30} we show the largest $\lambda=3.0$ that we could calculate in a reasonable time. Data still seems to be compatible with diffusion with small $D\approx 0.01$ for $U=0.1$. We note that even with large $\chi \sim 300$ an estimated error of the NESS current is still around $10-20\%$ for $L=64-160$, and therefore we can not unconditionally exclude a slight subdiffusion, $j \sim 1/L^\gamma$ with $\gamma \lesssim 1.1$ (for $U=0.1$). Because the slope of $j(L)$ though is decreasing with increasing $L$, and because the fitted $D\approx 0.01$ still agrees reasonably well with the exponential dependence on $\lambda$ obtained from the more precise data for smaller $\lambda$ (see Fig.~\ref{fig:D}) we judge that we still see diffusion. For $\lambda=2.5$ and $U=0.05$ (Fig.\ref{fig:lam30}), despite smaller $U=0.05$, we get NESS current precision of better than $10\%$ for $L \le 240$ and with $\chi < 200$, making us confident that we indeed see diffusion for those parameters.

In Fig.~\ref{fig:profil22} we show an example of steady state magnetization profiles. While for small $L$ one has a subdiffusive-like profile that is still influenced by a particular local potential values, asymptotically at large $L$ the profile gets linear (the amplitude of profile deviations from a linear one due to local potential also decreases) as expected for a diffusive system. In the unitary case one sees an excellent agreement with the error function profile at large times (Fig.~\ref{fig:sfig3}), however, there are deviations from this at short times, similar to what happens in the NESS setting for small system sizes. The reasoning is also similar as one needs to wait for the system to spread enough such that it is no longer dependent solely on the local potential values. 

Observing the convergence of the transferred magnetization in the unitary case with the bond dimension $\chi$, one sees (Fig.~\ref{fig:sfig1}) that it typically decreases, with the values differing by less than $5\%$ for $\chi \in [50-200]$ and not too large $\lambda$.

\subsection*{Single-particle analysis}

\begin{figure}[thb]
\centering
\includegraphics[width=0.77\linewidth]{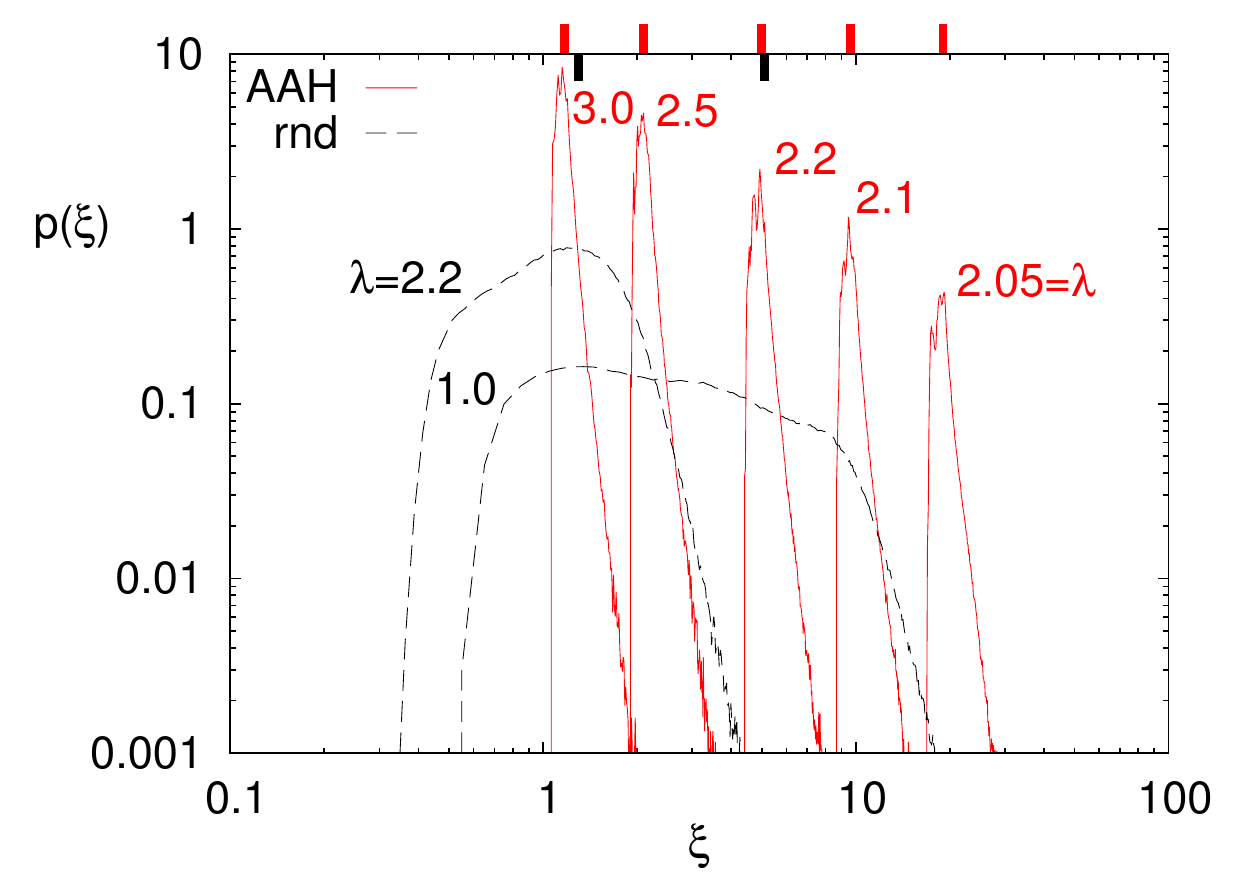}
\caption{Distribution of single-particle localization lengths $\xi$ for the AAH model (full red curves) and for the Anderson model (black dashed curves), and different $\lambda$. Short marks at the top axis denote the average $\xi$ for each respective data-set. In the Anderson (random phase) model the distributions are much wider. $L=10^6$.}
\label{fig:ksi}
\end{figure}
For single-particle analysis, that is for $U=0$, we diagonalize a long chain with $L=10^6$ in order to have sufficient statistics for various calculations. Because the eigenstates are exponentially localized we can split diagonalization of a long chain into separate diagonalizations of shorter overlapping segments (say of $\sim 10^4$ consecutive sites). From two overlapping segments we take a union of those eigenstates and eigenvalues that are far from the boundaries. Doing that we in the end get all the eigenstates of the long chain within the accuracy $\sim 10^{-10}$. We work in the fermionic language of spinless fermions $c_j$, in which the noninteracting $H$ is a quadratic function of fermionic operators,
\begin{equation}
  H=\vec{\mathbf{c}}^\dagger A \vec{\mathbf{c}},\quad \vec{\mathbf{c}}=(c_1,c_2,\ldots,c_L),
  \label{eq:A}
\end{equation}
where $A$ is a tridiagonal matrix with nonzero elements $A_{j,j}=-2h_j$, $A_{j,j+1}=A_{j+1,j}=2$. The eigenvalues of $A$ give us the single-particle eigenenergies $E_k$, while the eigenstates $\psi_k$, packed into a unitary matrix $U_{j,k}:=[\psi_k]_j$, determine a canonical transformation into new fermionic operators $f_j$, $\vec{\mathbf{c}}=U \vec{\mathbf{f}}$, such that we can write the Hamiltonian as $H=\sum_j E_j f_j^\dagger f_j$. 

As an example we show in Fig.~\ref{fig:spekterAAH} all single-particle eigenenergies $E_k$ for the AAH chain of length $L=10^6$. While we would have a finite number of bands for a periodic potential, i.e., rational $\beta$, for an irrational $\beta$ (we always use $\beta=(\sqrt{5}-1)/2$) one has an infinite sequence of ''sub-bands'' resulting in a Cantor set.

Next we determine the distribution of localization lengths. To that end we fit each eigenstate by $|[\psi_k]_j|^2 \sim \exp{(-|j-c_k|/\xi_k)}$, where $c_k=\sum_j j |[\psi_k]_j|^2$, obtaining its localization length $\xi_k$. In Fig.~\ref{fig:ksi} we show the distribution $p(\xi)$ of localization lengths, its average being $\ave{\xi}\approx \frac{1}{2\log{(\lambda/2)}}$ for the AAH model. We can see that the distribution of localization lengths is much wider for the Anderson model than in the quasiperiodic AAH model. This is reflected also in the full many-body calculations -- fluctuations between NESS currents for different realizations are much larger in the Anderson model than in the AAH model, see Fig.~\ref{fig:cmp}a.

\begin{figure*}[tb!]
\centerline{\includegraphics[width=0.4\linewidth]{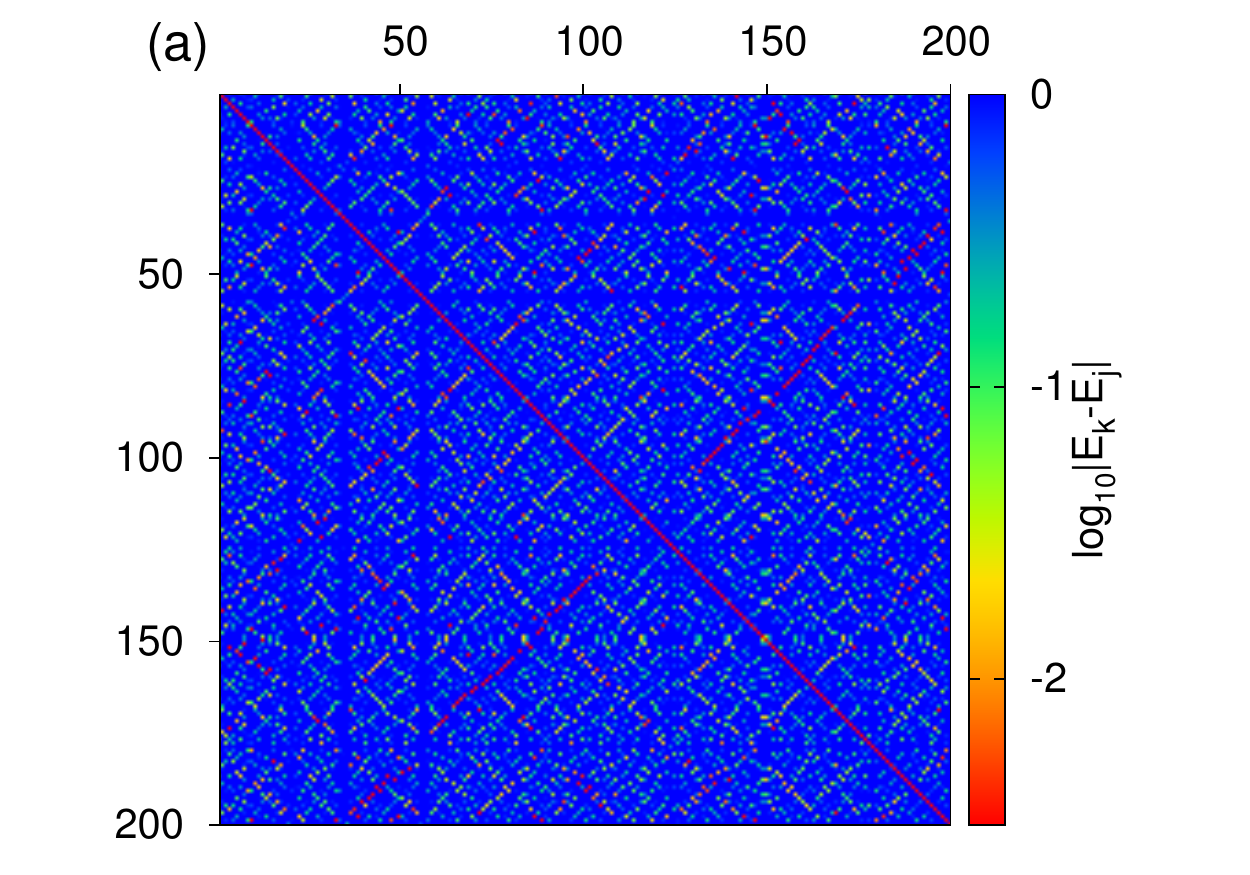}\includegraphics[width=0.4\linewidth]{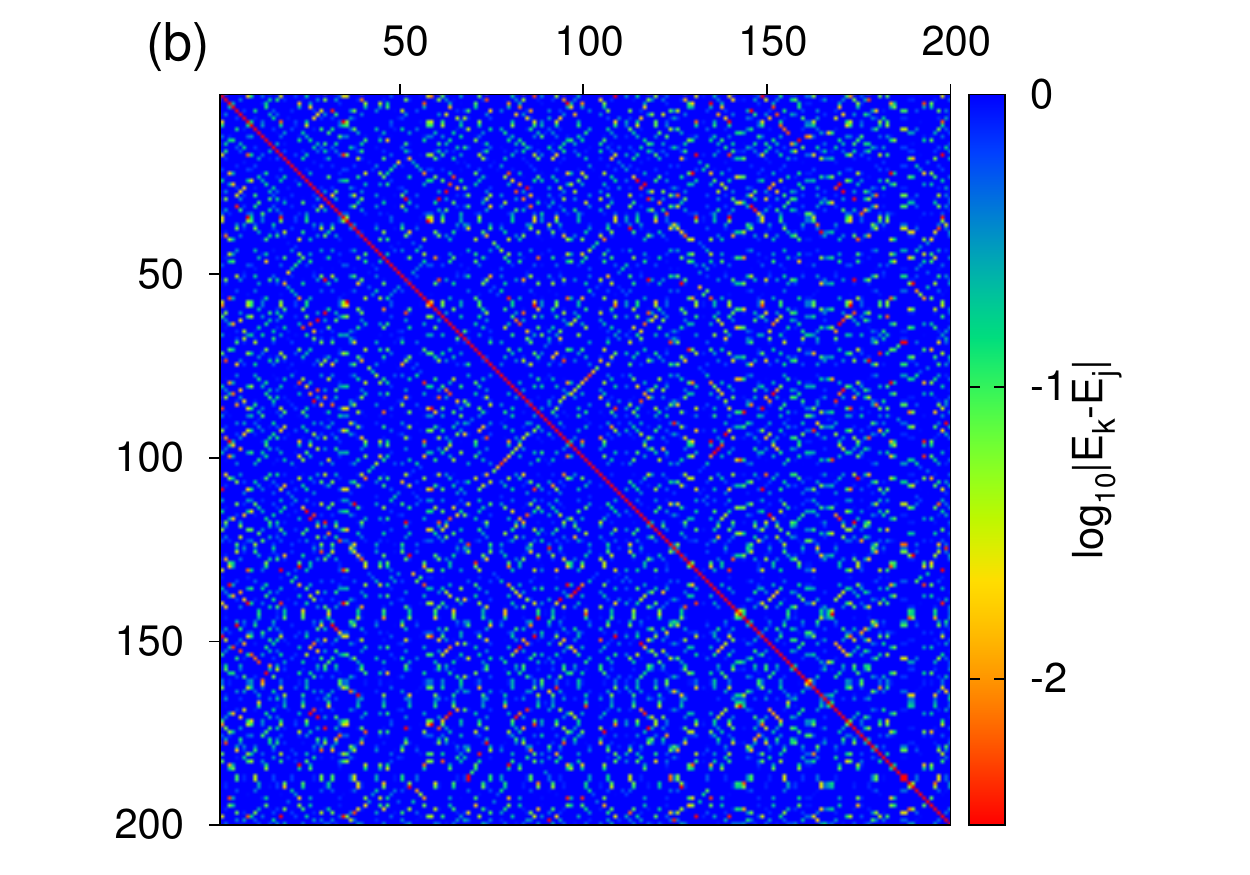}}
\centerline{\includegraphics[width=0.4\linewidth]{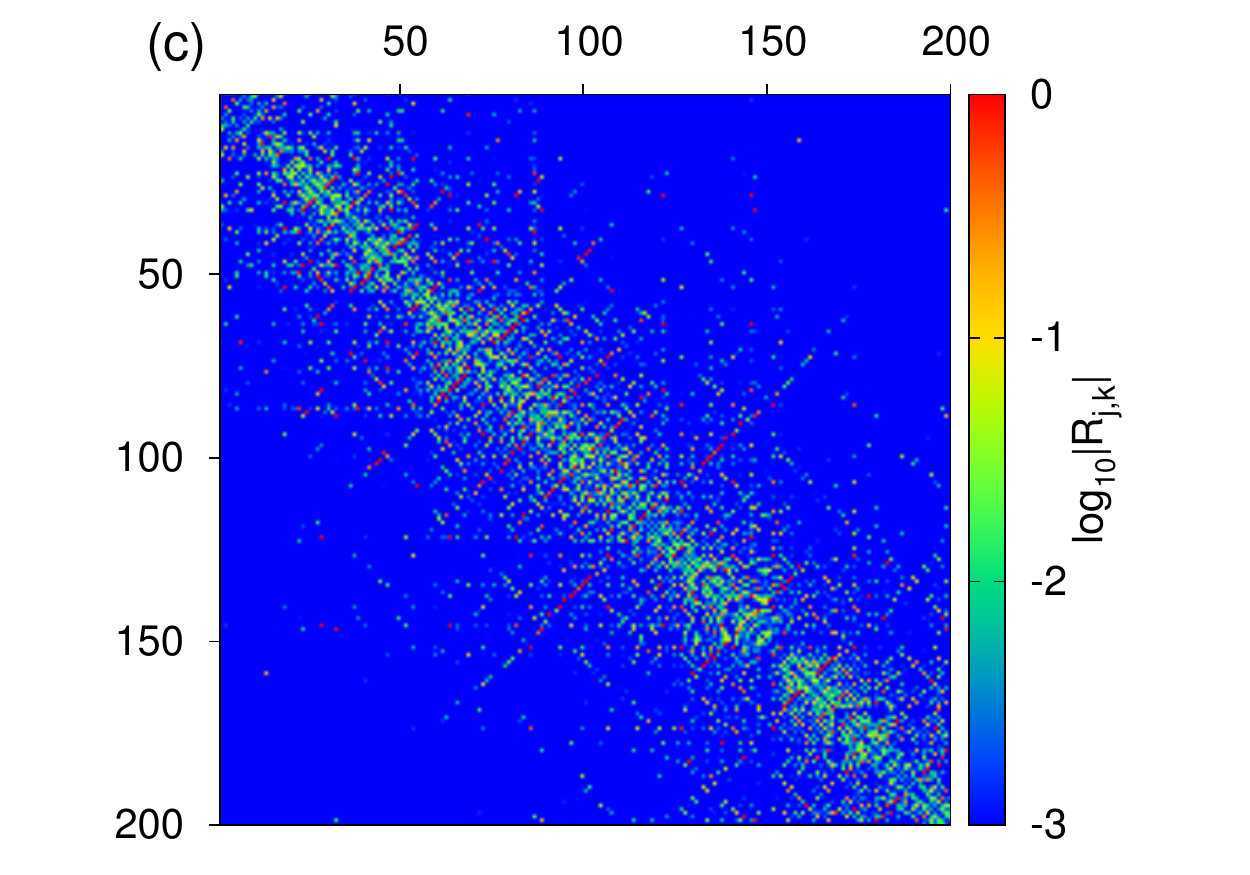}\includegraphics[width=0.4\linewidth]{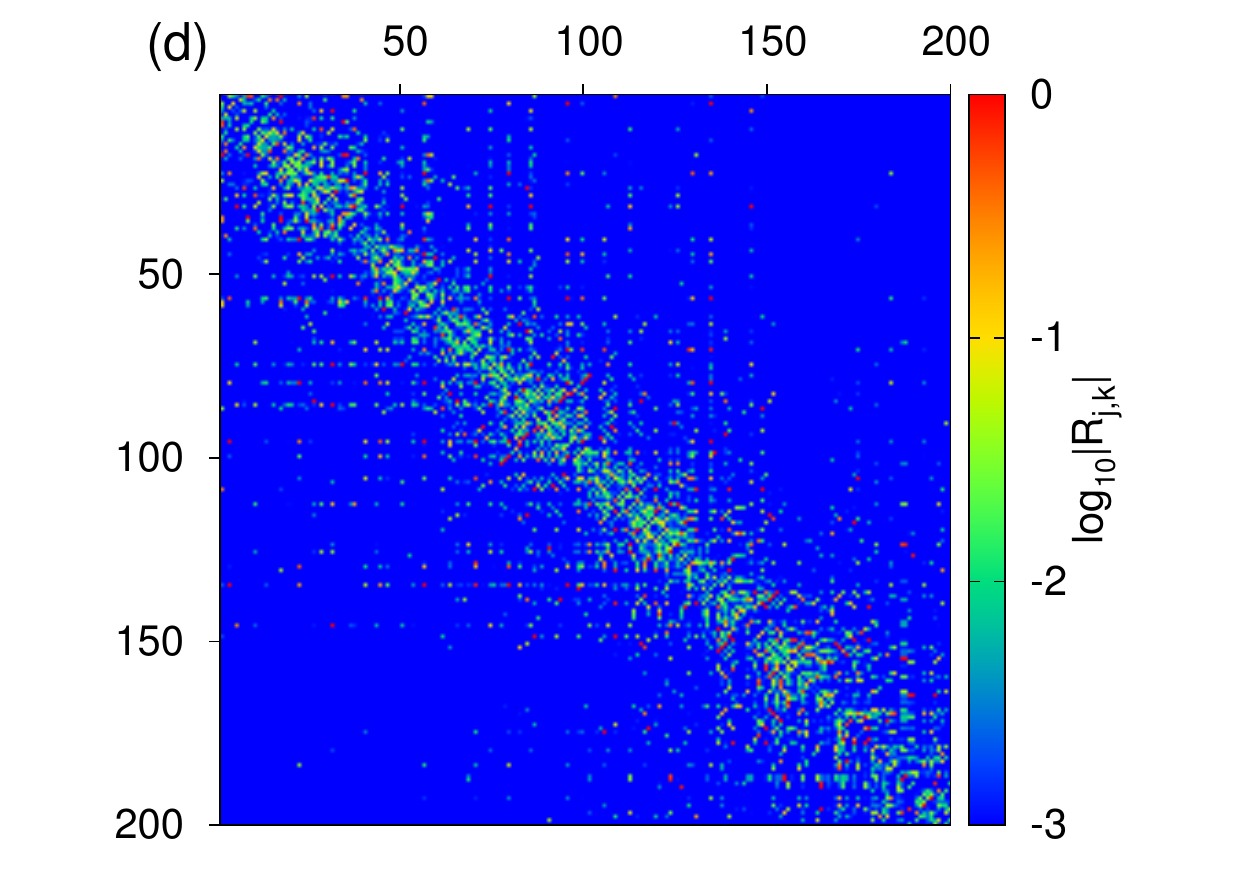}}
\caption{Noninteracting AAH model ($\lambda=2.2$) and single-particle resonances when randomizing the phase on some sites. Left column (frames (a) and (c)) is for $\delta=0.01$, right column (frames (b) and (d)) for $\delta=0.05$. Top row, frames (a) and (b), show resonances in the single-particle energies $E_k$, while (c) and (d) show large resonance factors $R_{j,k}$, Eq.~\ref{eq:R}. Eigenstates are ordered (matrix index $j,k \in [1,200]$) according to their location along the chain and we show $200$ consecutive eigenstates from the bulk. Resonances in all plots are visible as red (bright) dots.}
\label{fig:reskick}
\end{figure*}

Behavior for small interactions $U$ can be in principle obtained from the perturbation series in $V$. In fermionic language we have (up to irrelevant constants)
\begin{equation}
  V=\sum_j n_j n_{j+1}, \qquad n_j=c_j^\dagger c_j.
\end{equation}
Because $V$ involves interaction between 2 fermions all matrix elements that would change only one single-particle state are zero, $\bracket{\alpha}{V}{\beta}=0$, where for brevity we denote the occupied single-particle states $\psi_\alpha$ with energy $E_\alpha$ simply by their integer ``location'' index $\alpha$ (that are ordered according to their center-of-mass location $c_\alpha$). Nonzero though are 2-particle matrix elements,
\begin{eqnarray}
&&  \bracket{\alpha_1 \alpha_2}{V}{\beta_1 \beta_2}=\sum_j U^*_{j,\alpha_1} U_{j,\beta_1} U^*_{j+1,\alpha_2} U_{j+1,\beta_2}+\nonumber \\
  &&  +U^*_{j,\alpha_2} U_{j,\beta_2} U^*_{j+1,\alpha_1} U_{j+1,\beta_1}-U^*_{j,\alpha_2} U_{j,\beta_1} U^*_{j+1,\alpha_1} U_{j+1,\beta_2}-\nonumber \\
&&  -U^*_{j,\alpha_1} U_{j,\beta_2} U^*_{j+1,\alpha_2} U_{j+1,\beta_1}.
\end{eqnarray}
Due to localization of single-particle eigenstates these matrix elements get exponentially suppressed as soon as indices $\alpha_j$ and $\beta_j$ are far apart. Therefore, the largest non-diagonal elements are those where one fermion does not jump, that is
\begin{equation}
  \bracket{\alpha_1,\alpha_2}{V}{\alpha_1,\beta_2}.
  \label{eq:V}
  \end{equation}
They describe density-mediated jumps of a single fermion. Whether the two states $\alpha_2$ and $\beta_2$ get mixed (hybridized) by the perturbation depends on the energy difference $E_{\alpha_2}-E_{\beta_2}$. If the matrix element is comparable or larger than the energy miss-match the states get mixed, otherwise not. Therefore, in Fig.~\ref{fig:res} we analyze the ratio of the two, Eq.\ref{eq:R}. We have checked that a similar but less strong pattern is obtained also if $\beta_1$ is not exactly equal to $\alpha_1$ (as in Eq.\ref{eq:V}), e.g., for $\beta_1=\alpha_1 \pm p$ with small $p$.

\subsection*{Random phase-kicks}

\begin{figure}[thb]
\centering
\includegraphics[width=0.77\linewidth]{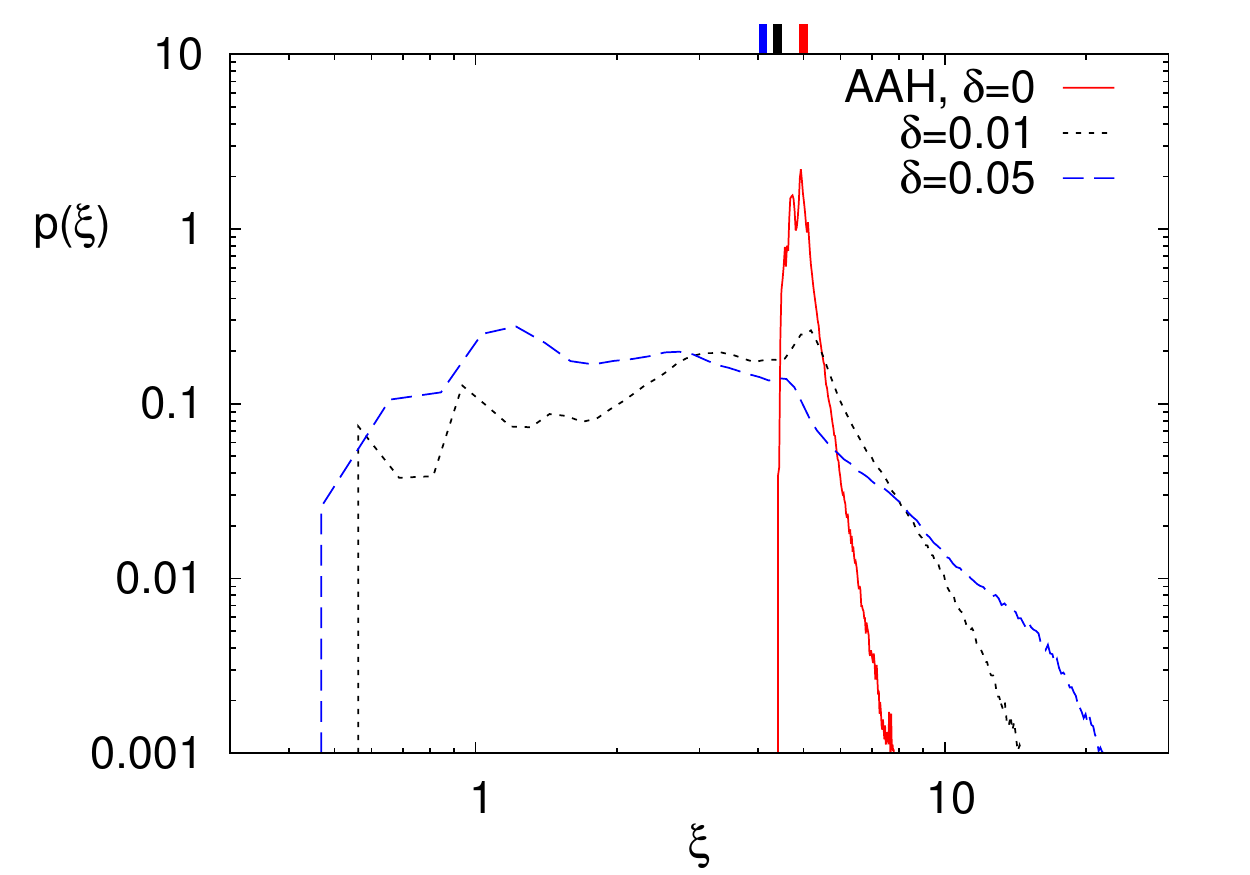}
\caption{Distribution of single-particle localization lengths $\xi$ for the AAH model (full red curve), as well as for the AAH model with random phases for $\delta=0.01$ and $\delta=0.05$. As soon as $\delta \neq 0$ the distribution gets wider. All is for $\lambda=2.2$ and $L=10^6$. Short marks at the top axis denote the average $\xi$ for each respective data-set.}
\label{fig:ksikick}
\end{figure}

We have demonstrated in the main text that the transport in the AAH model can completely change once one breaks the long-range phase coherence of the potential. Specifically, we randomize the phase at each site $j$ that satisfies the condition $1-\cos{(2\pi\beta j+\phi)}<\delta$ for some small $\delta$. We have seen  in Fig.~\ref{fig:kick}b that the distribution of nearest-resonance distances gets similar to the one for the Anderson model already for very small $\delta$. In Fig.~\ref{fig:ksikick} we show that similar thing happens also for the distribution of localization lengths. As soon as $\delta >0$ the distribution gets wide, much like in the Anderson model.
\newpage 
In Fig.~\ref{fig:reskick} we show a spatial resonance structure for the AAH model with randomized phases, similar to the AAH model data in Fig.~\ref{fig:res}. We can see that already for $\delta=0.01$ the gross resonance structure is washed out, and we conjecture that the transport is subdiffusive in the TDL for any nonzero $\delta$.

\end{document}